\documentclass{vgtc}                          

\ifpdf
  \pdfoutput=1\relax                   
  \usepackage{graphicx}                
  \DeclareGraphicsExtensions{.pdf,.png,.jpg,.jpeg} 
\else
  \usepackage{graphicx}                
  \DeclareGraphicsExtensions{.eps}     
\fi%

\graphicspath{{figures/}{pictures/}{images/}{./}} 

\usepackage{algorithm}
\usepackage{algpseudocode}
\usepackage{amsmath}

\usepackage{microtype}                 
\PassOptionsToPackage{warn}{textcomp}  
\usepackage{textcomp}                  
\usepackage{mathptmx}                  
\usepackage{amssymb}                   %
\usepackage{times}                     
\usepackage{cite}                      
\usepackage{tabu}                      
\usepackage{booktabs}                  
\usepackage{subfig}

\usepackage{tikz}
\usepackage{multirow}
\onlineid{0}

\vgtccategory{Research}

\vgtcinsertpkg

\usepackage{color}

\newenvironment{DIFnomarkup}{}{}

\title{Accelerating Force-Directed Graph Drawing with RT Cores}

\author{Stefan Zellmann\thanks{e-mail: zellmann@uni-koeln.de}\\ %
        \scriptsize University of Cologne %
\and Martin Weier\thanks{e-mail: martin.weier@h-brs.de}\\ %
     \scriptsize Hochschule Bonn-Rhein-Sieg %
\and Ingo Wald\thanks{e-mail: iwald@nvidia.com}\\ %
     \parbox{1.4in}{\scriptsize \centering NVIDIA}}

\teaser{
\vspace{-2em}

  \setlength{\fboxrule}{0pt}
  \fbox{\begin{tikzpicture}
    \node[anchor=south west,inner sep=0] (image) at (0,0)
        { \includegraphics[width=1.45in]{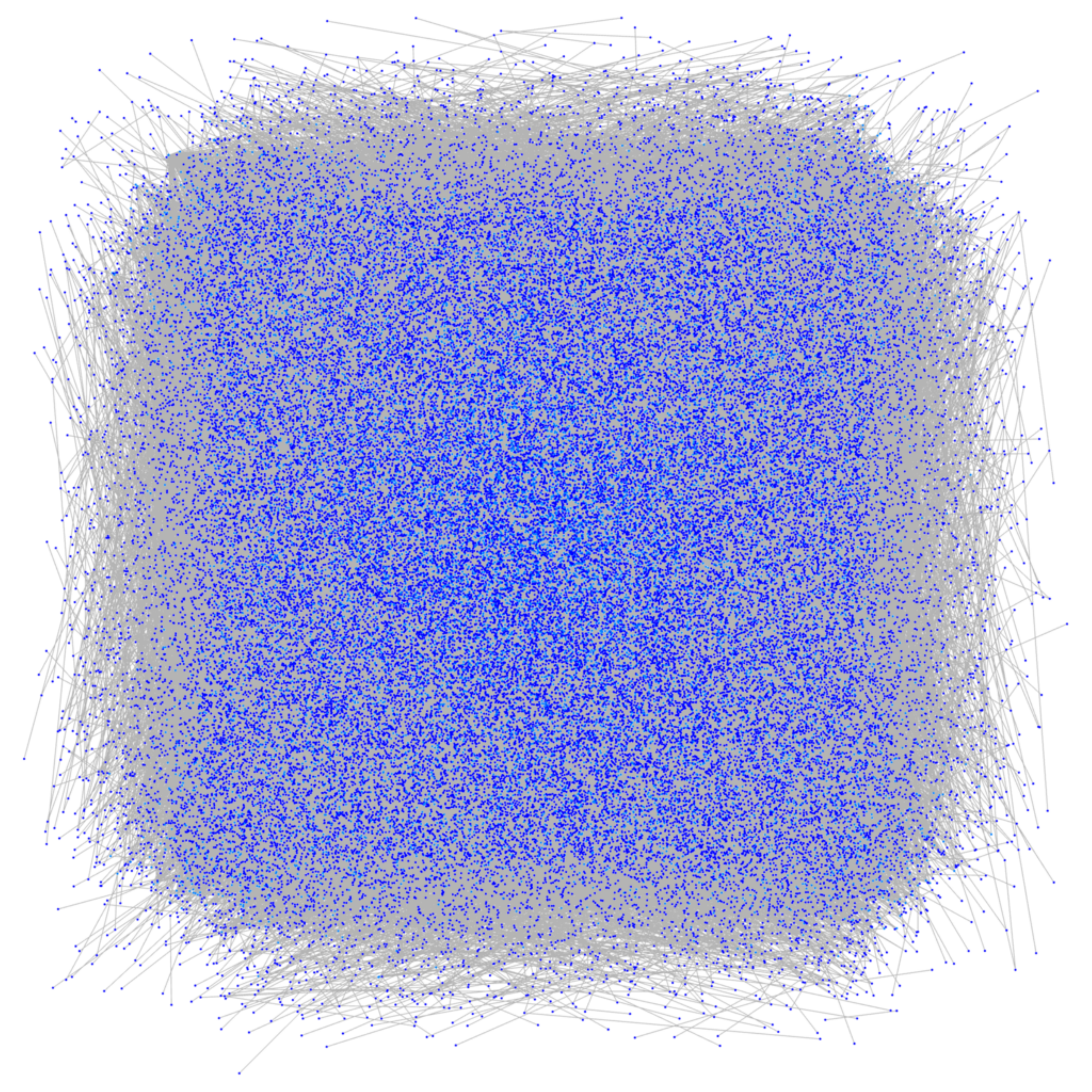}};
    \begin{scope}[x={(image.south east)},y={(image.north west)}]
      \draw[red,thick,rounded corners] (0.39,0.6) rectangle ++(0.17,0.17);
      \draw[red,thick,rounded corners] (0.49,0.32) rectangle ++(0.17,0.17);
    \end{scope}
  \end{tikzpicture}}
  \fbox{\begin{tikzpicture}
    \node[anchor=south west,inner sep=0] (image) at (0,0)
        { \includegraphics[width=1.45in]{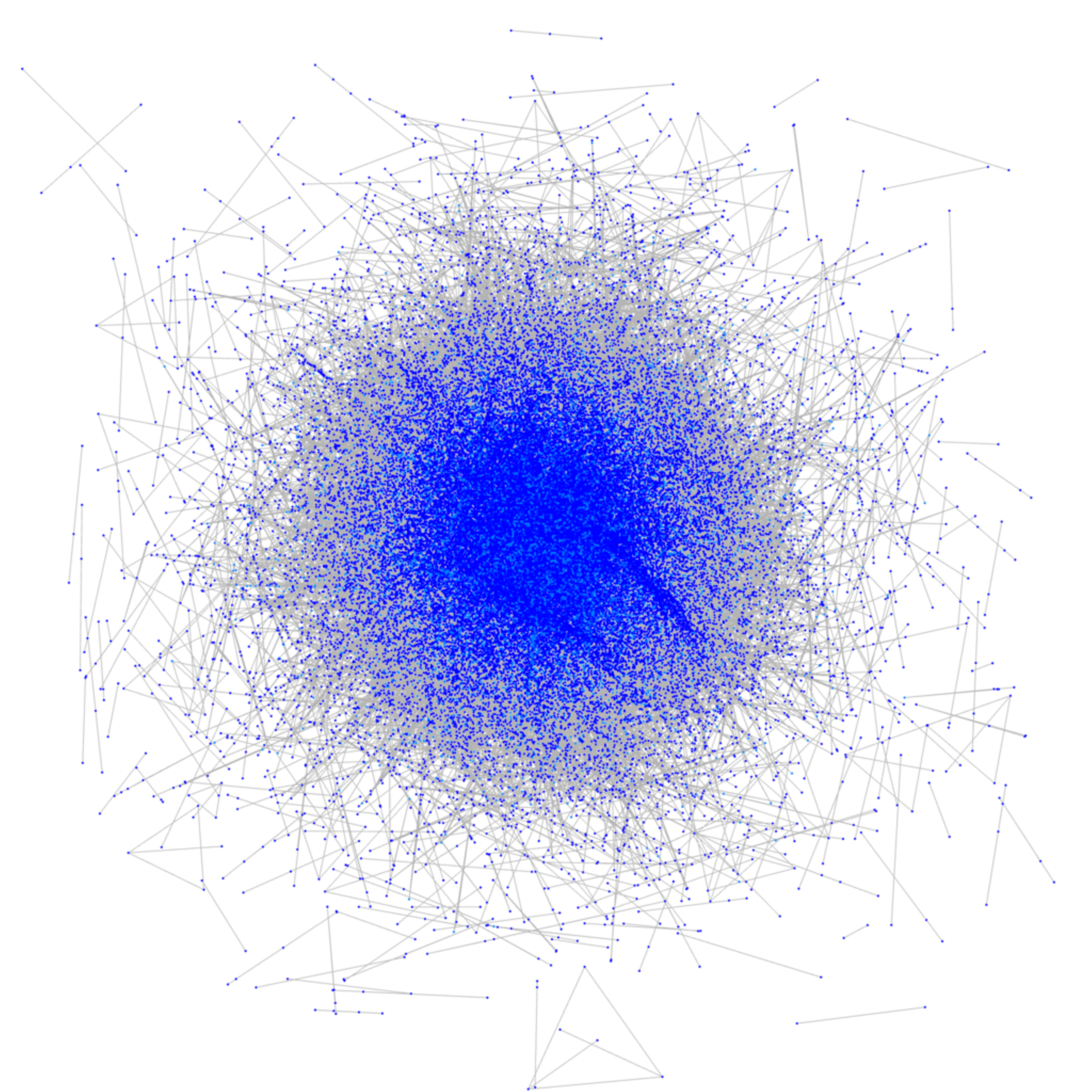}};
    \begin{scope}[x={(image.south east)},y={(image.north west)}]
      \draw[red,thick,rounded corners] (0.39,0.6) rectangle ++(0.17,0.17);
      \draw[red,thick,rounded corners] (0.49,0.32) rectangle ++(0.17,0.17);
    \end{scope}
  \end{tikzpicture}}
  \fbox{\begin{tikzpicture}
    \node[anchor=south west,inner sep=0] (image) at (0,0)
        { \includegraphics[width=1.45in]{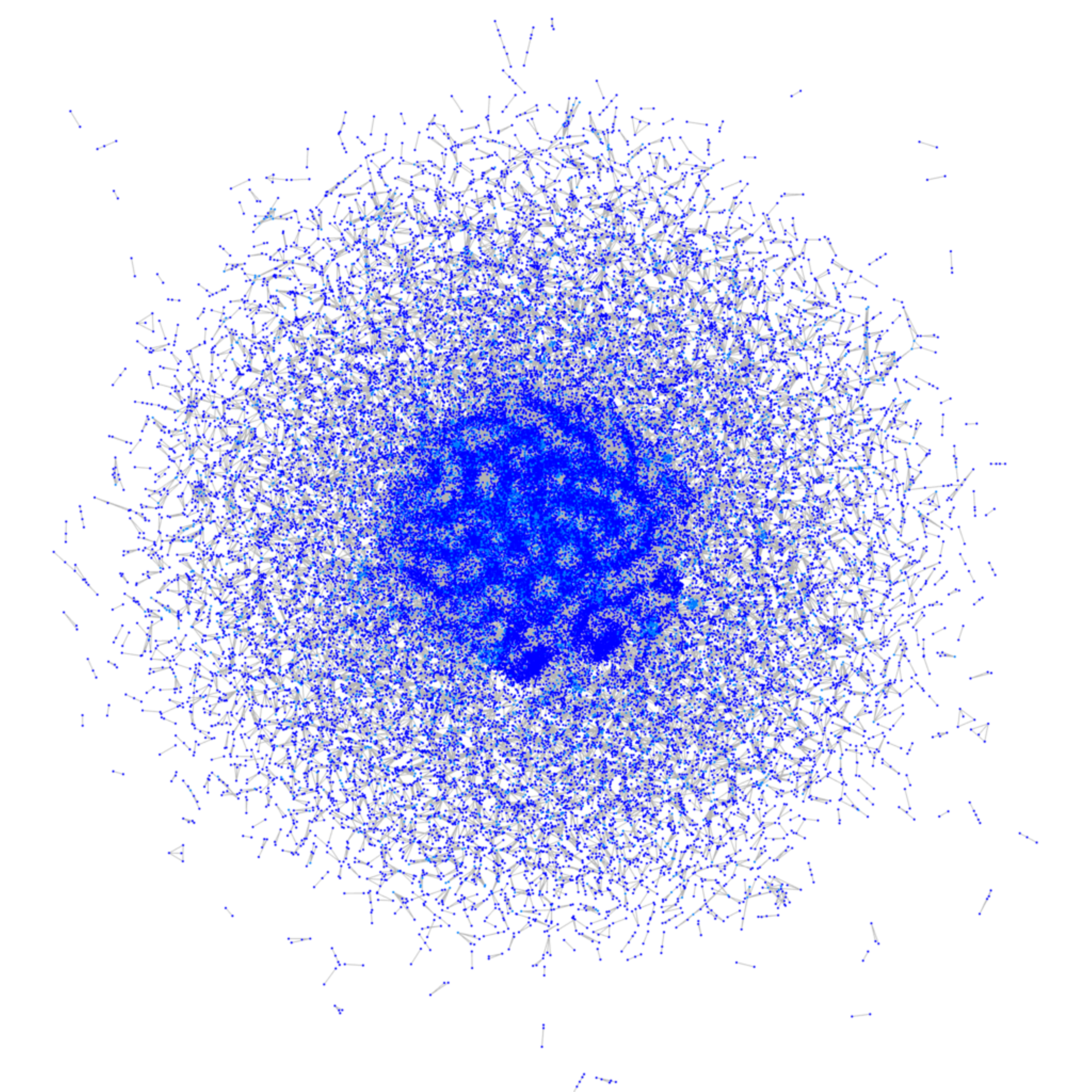}};
    \begin{scope}[x={(image.south east)},y={(image.north west)}]
      \draw[red,thick,rounded corners] (0.39,0.6) rectangle ++(0.17,0.17);
      \draw[red,thick,rounded corners] (0.49,0.32) rectangle ++(0.17,0.17);
    \end{scope}
  \end{tikzpicture}}
  \fbox{\begin{tikzpicture}
    \node[anchor=south west,inner sep=0] (image) at (0,0)
        { \includegraphics[width=1.45in]{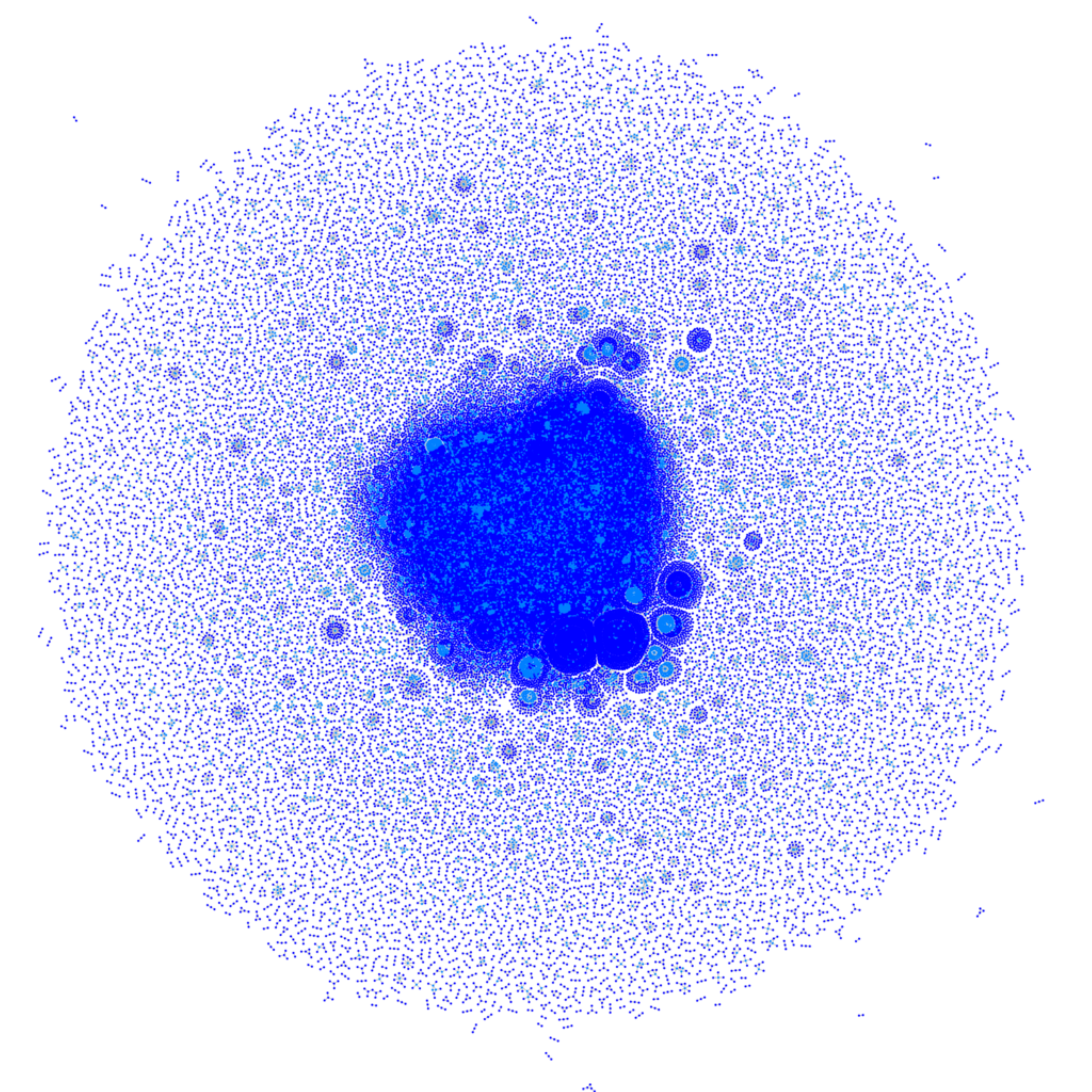}};
    \begin{scope}[x={(image.south east)},y={(image.north west)}]
      \draw[red,thick,rounded corners] (0.39,0.6) rectangle ++(0.17,0.17);
      \draw[red,thick,rounded corners] (0.49,0.32) rectangle ++(0.17,0.17);
    \end{scope}
  \end{tikzpicture}}

    \vspace{-5pt}

  \fbox{\begin{tikzpicture}
    \node[anchor=south west,inner sep=0] (image) at (0,0)
        { \includegraphics[width=0.665in]{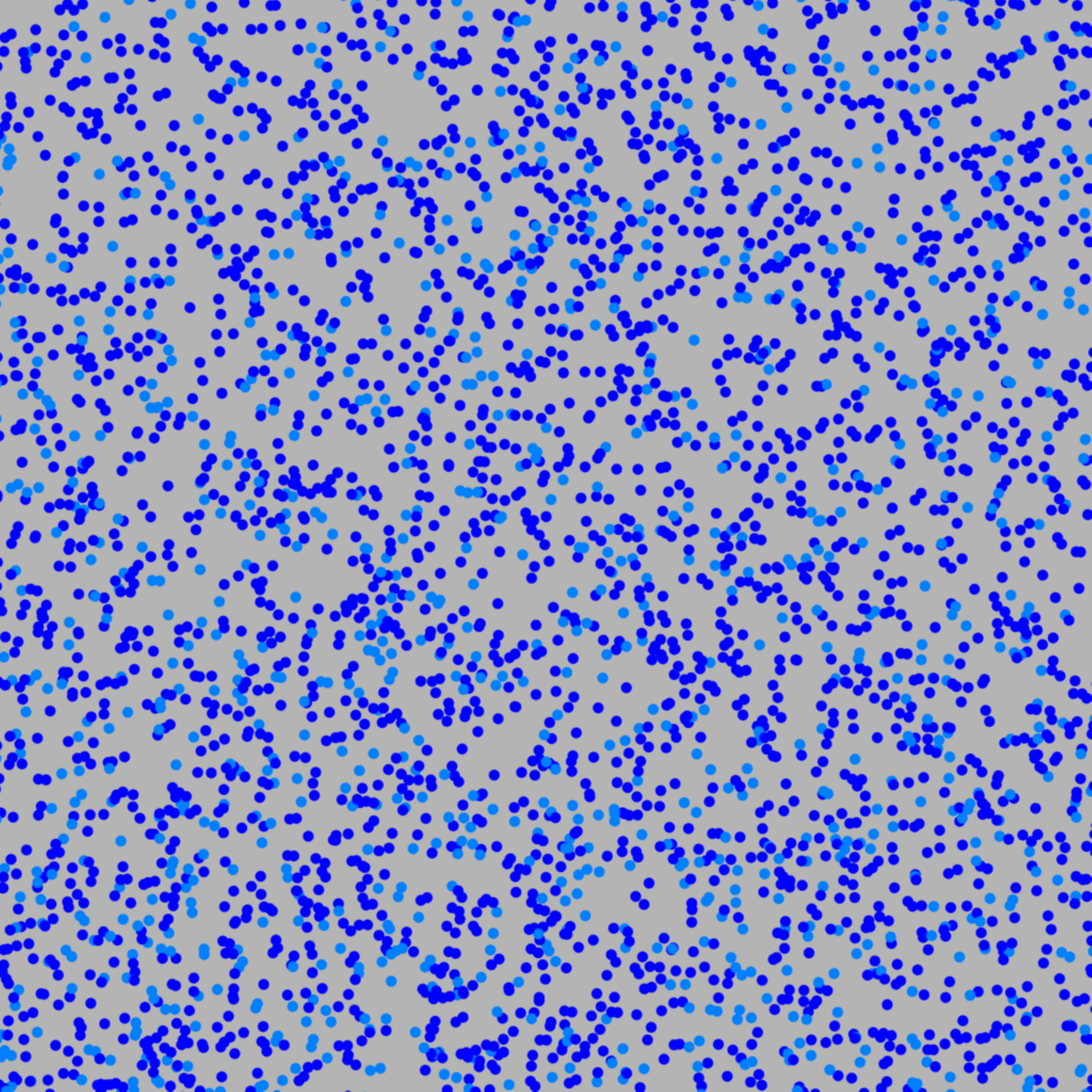}};
  \end{tikzpicture}}
  \fbox{\begin{tikzpicture}
    \node[anchor=south west,inner sep=0] (image) at (0,0)
        { \includegraphics[width=0.665in]{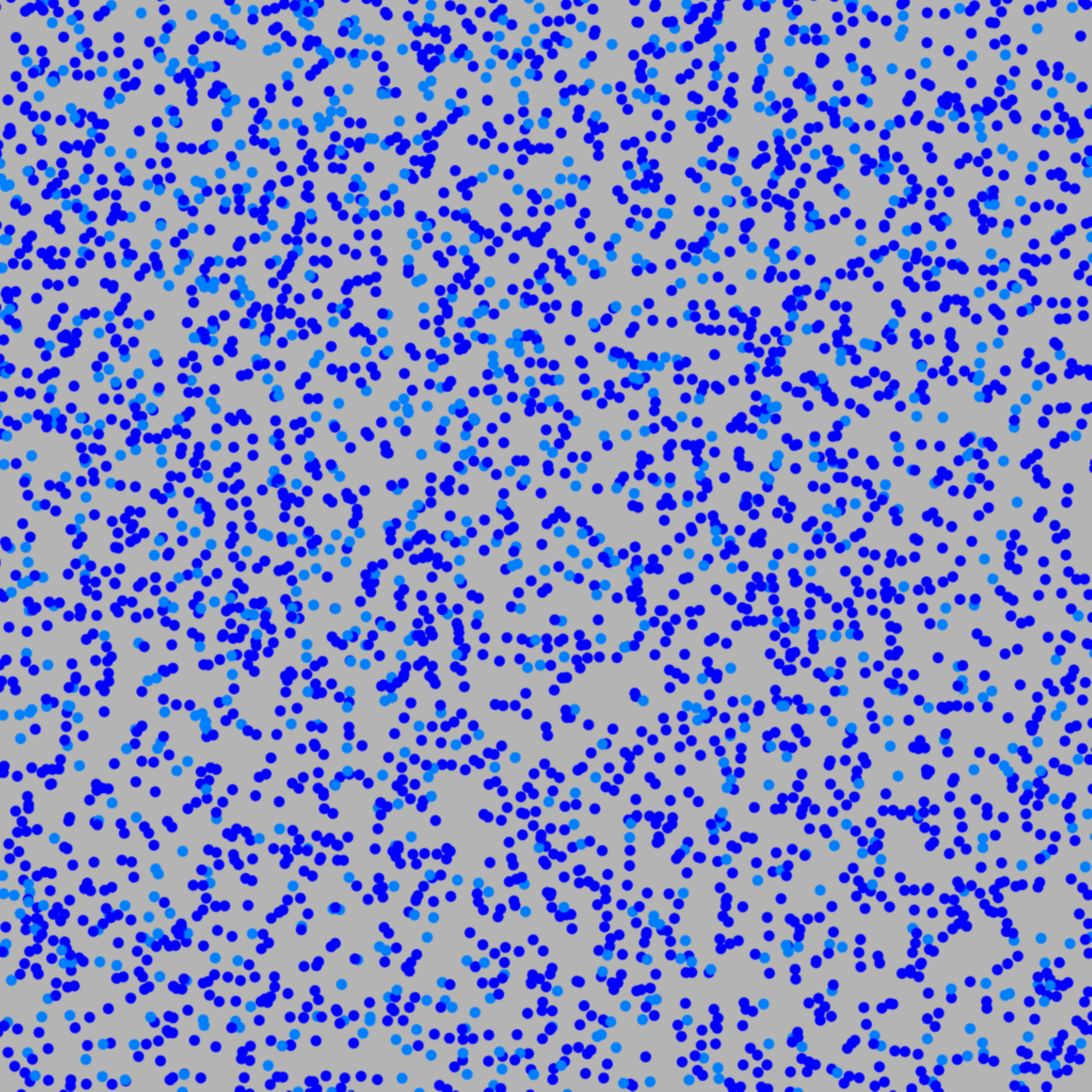}};
  \end{tikzpicture}}
  \fbox{\begin{tikzpicture}
    \node[anchor=south west,inner sep=0] (image) at (0,0)
        { \includegraphics[width=0.665in]{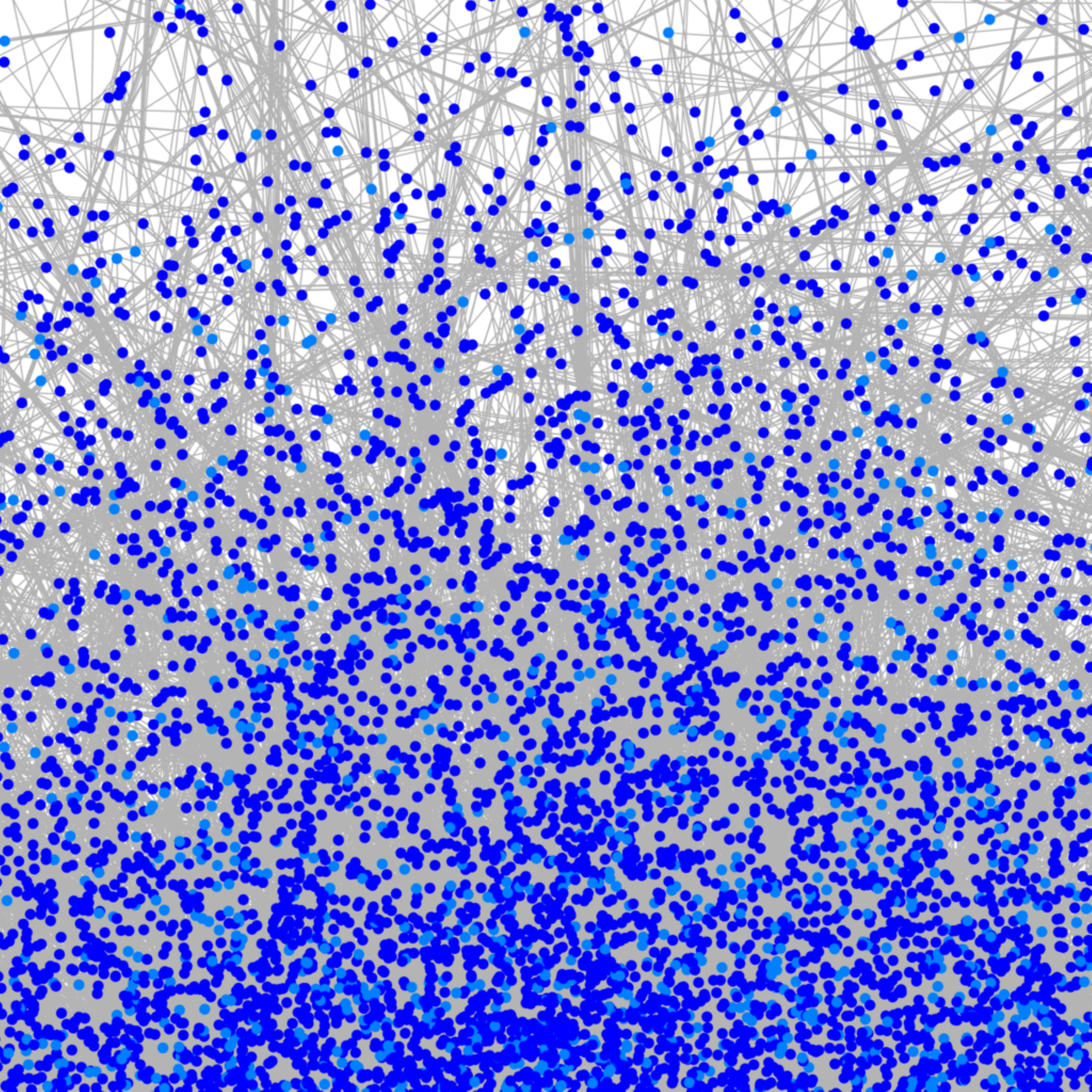}};
  \end{tikzpicture}}
  \fbox{\begin{tikzpicture}
    \node[anchor=south west,inner sep=0] (image) at (0,0)
        { \includegraphics[width=0.665in]{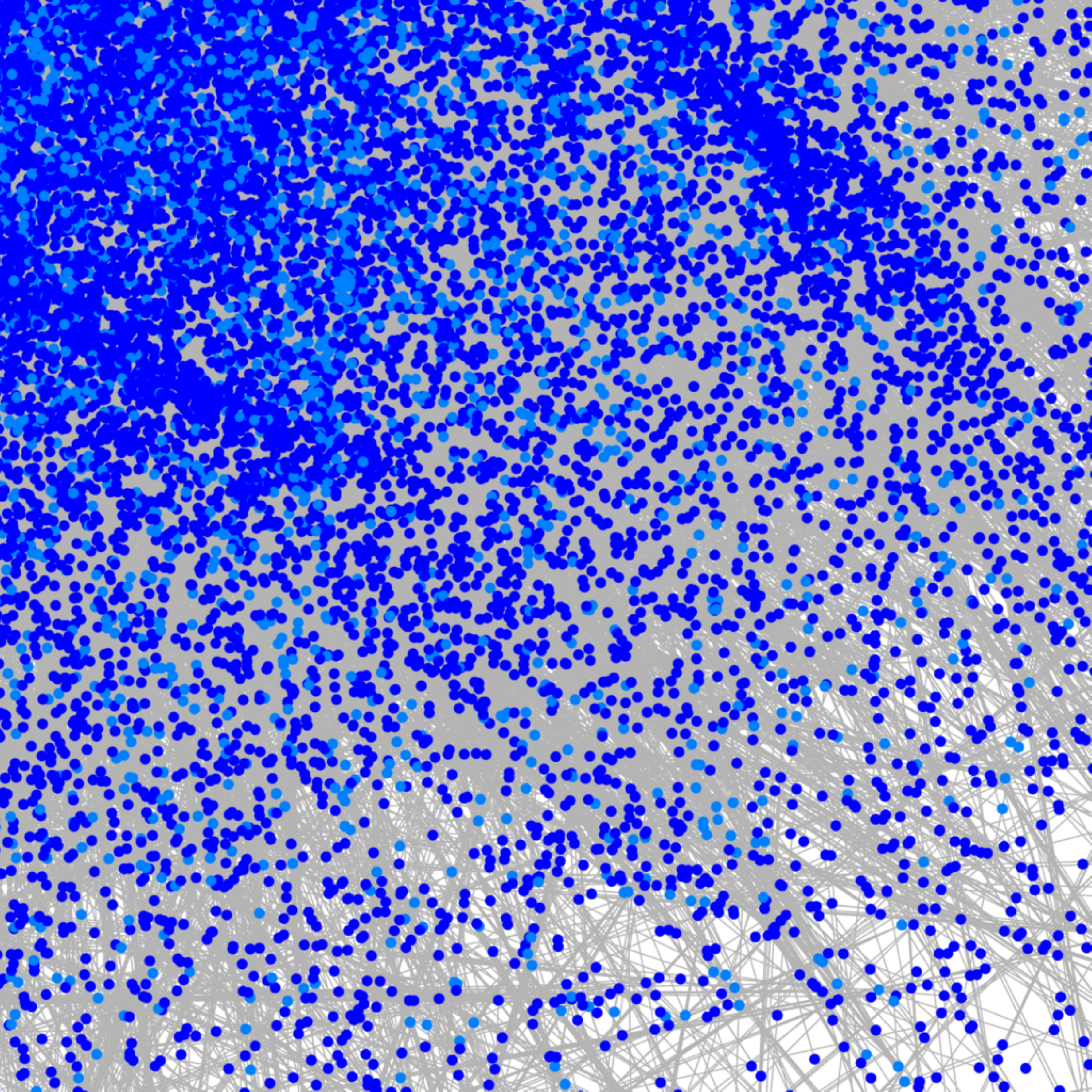}};
  \end{tikzpicture}}
  \fbox{\begin{tikzpicture}
    \node[anchor=south west,inner sep=0] (image) at (0,0)
        { \includegraphics[width=0.665in]{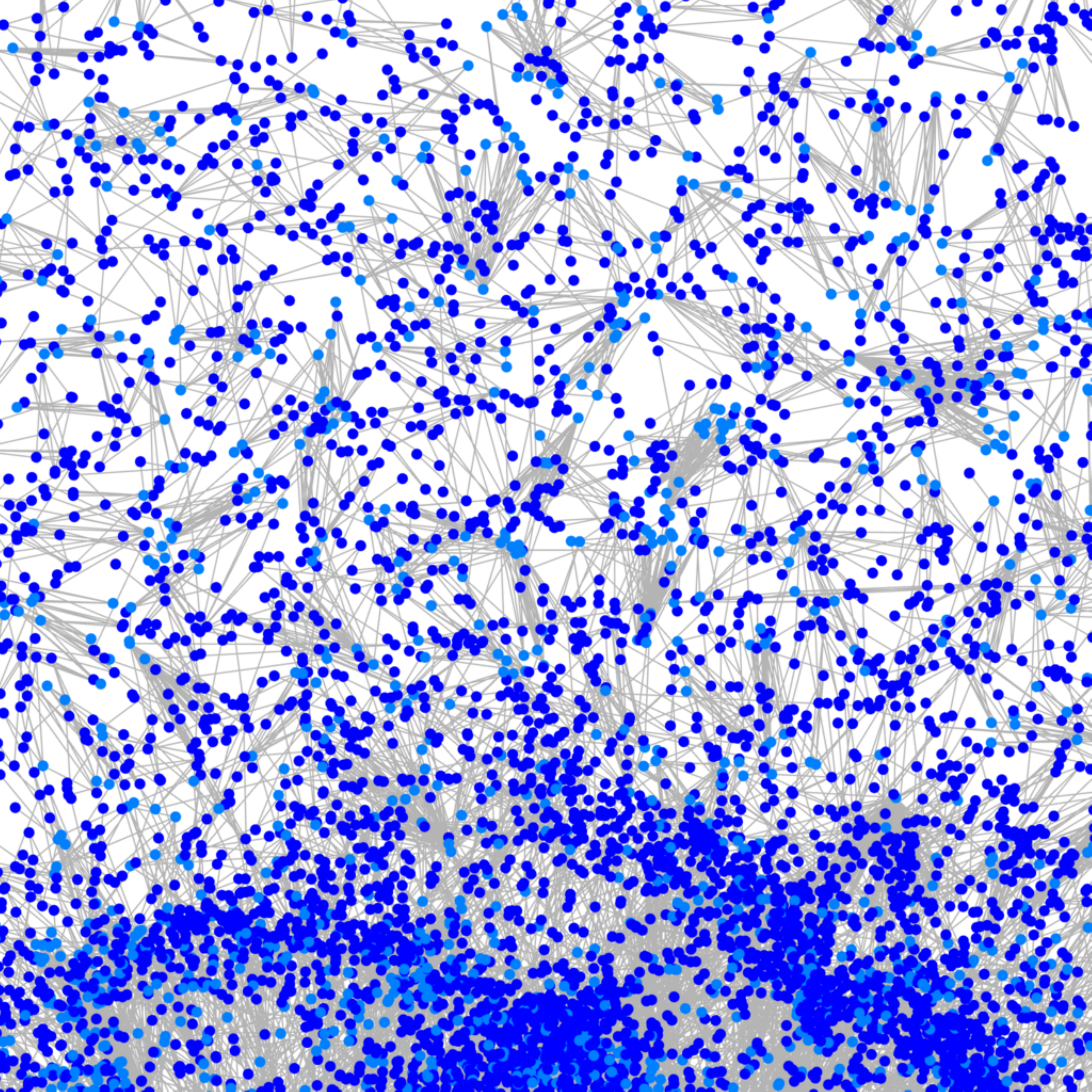}};
  \end{tikzpicture}}
  \fbox{\begin{tikzpicture}
    \node[anchor=south west,inner sep=0] (image) at (0,0)
        { \includegraphics[width=0.665in]{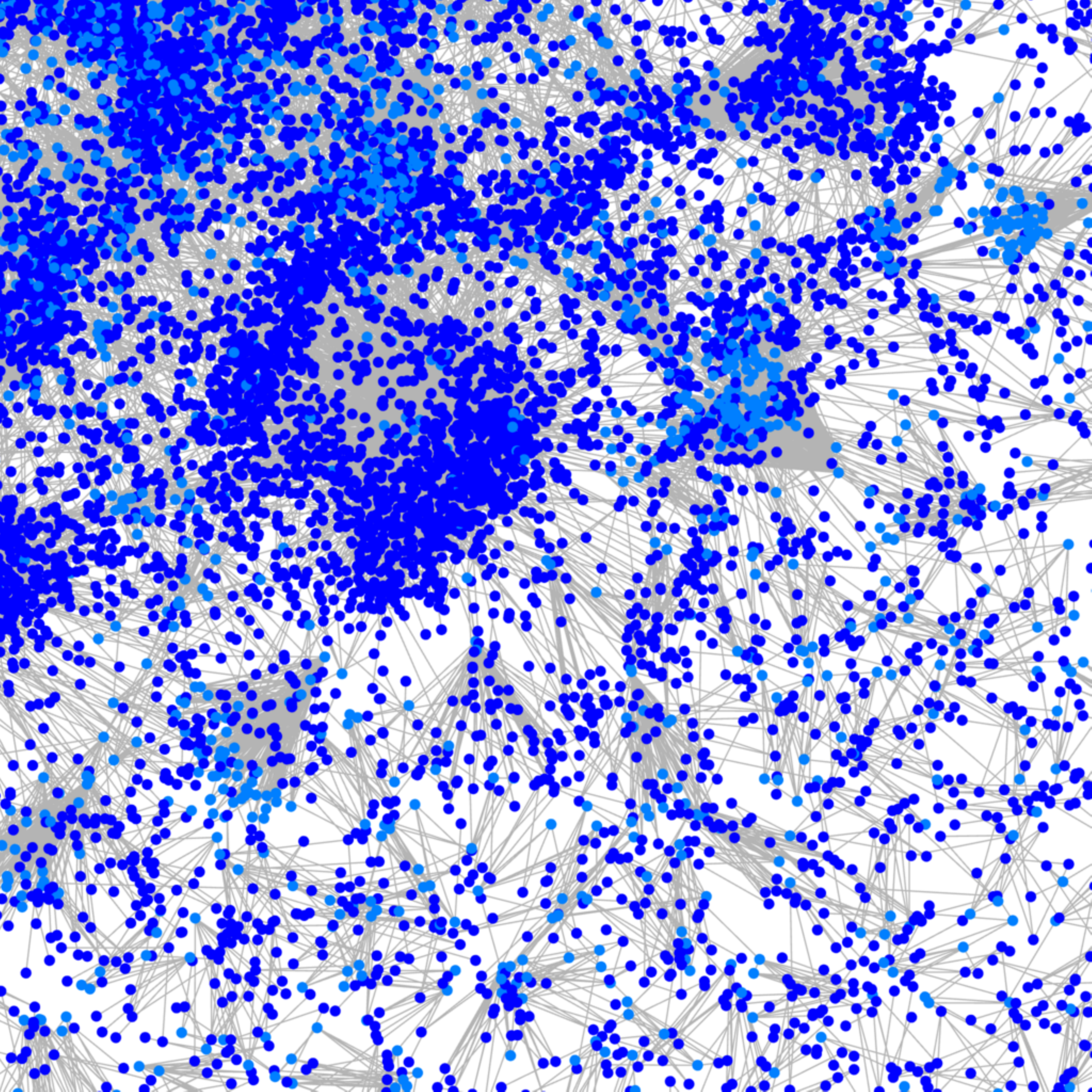}};
  \end{tikzpicture}}
  \fbox{\begin{tikzpicture}
    \node[anchor=south west,inner sep=0] (image) at (0,0)
        { \includegraphics[width=0.665in]{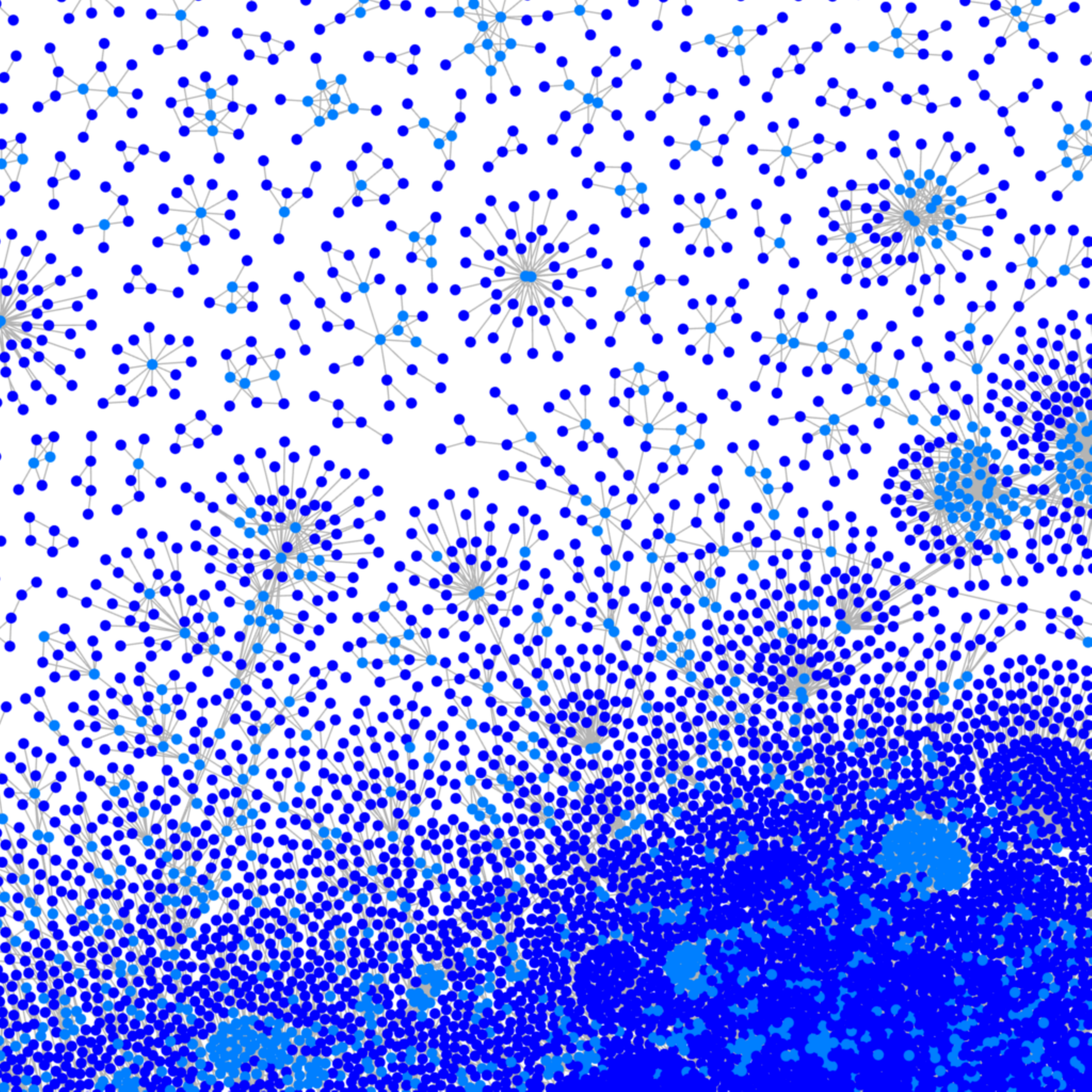}};
  \end{tikzpicture}}
  \fbox{\begin{tikzpicture}
    \node[anchor=south west,inner sep=0] (image) at (0,0)
        { \includegraphics[width=0.665in]{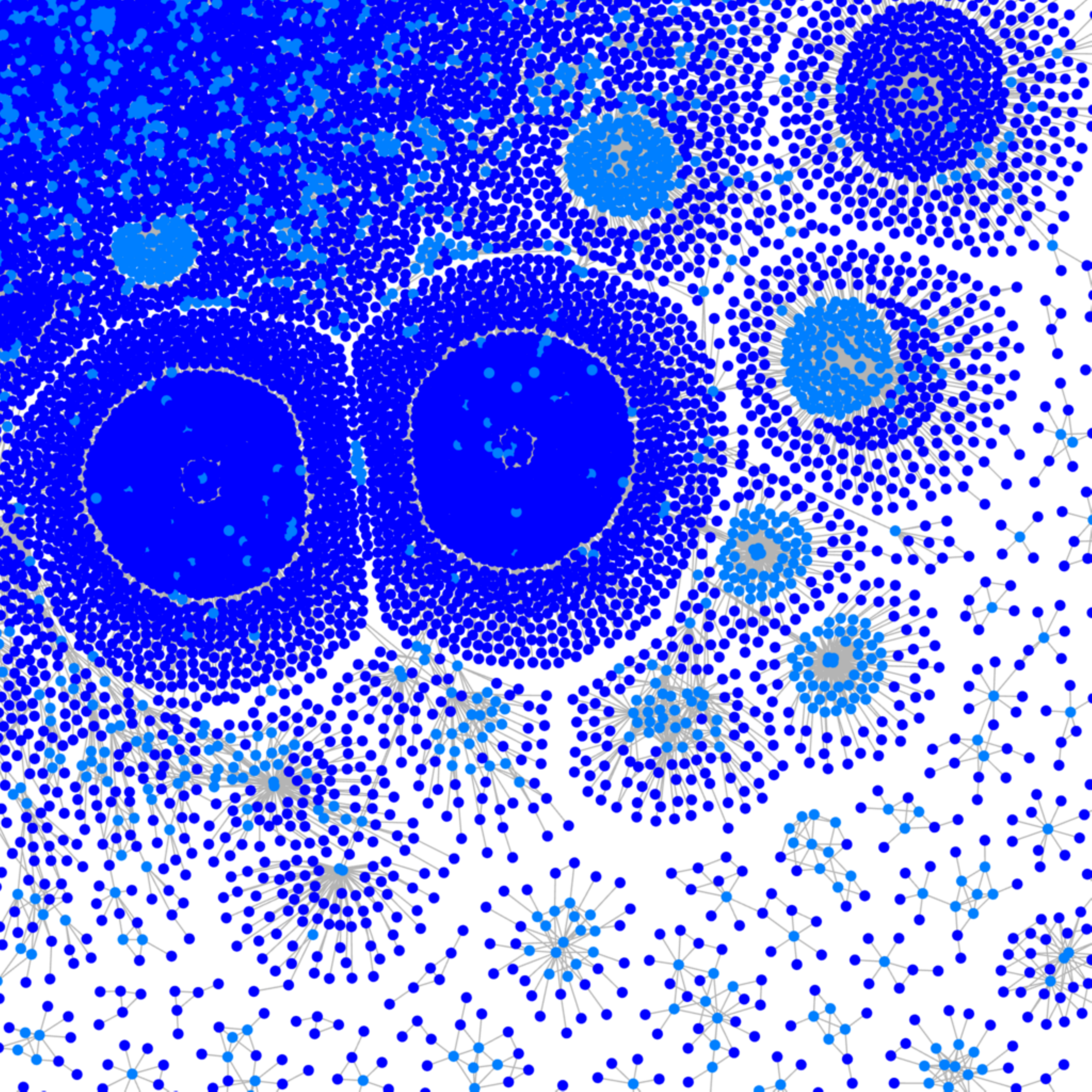}};
  \end{tikzpicture}}

  \vspace{-1em}
  \caption{
  Drawing a Twitter feed graph (68K vertices, 101K edges) with a force-directed
  algorithm using RT cores. The images show the results after $N=1$ (left), $N=100$ (second from left),
  $N=2{,}000$ (second from right), and $N=12{,}000$ (right) iterations.
  We can generate these layouts in $0.003$, $0.43$, $7.35$, and $39.3$
  seconds, and outperform a typical CUDA software implementation by $10.2\times$,
  $7.44\times$, $9.6\times$, and $10.9\times$, respectively.
  \label{fig:teaser}
  \vspace{-1em}
  }
}

\abstract{
Graph drawing with spring embedders employs a $V \times V$ computation phase
over the graph's vertex set to compute repulsive forces. Here, the efficacy of
forces diminishes with distance: a vertex can effectively only influence other
vertices in a certain radius around its position. Therefore, the algorithm
lends itself to an implementation using search data structures to reduce the
runtime complexity. NVIDIA RT cores implement hierarchical tree traversal in
hardware. We show how to map the problem of finding graph layouts with
force-directed methods to a ray tracing problem that can subsequently be
implemented with dedicated ray tracing hardware. With that, we observe speedups
of $4\times$ to $13\times$ over a CUDA software implementation.
} 

\CCScatlist{
  \CCScatTwelve{Human-centered computing}{Visu\-al\-iza\-tion}{Visu\-al\-iza\-tion techniques}{Graph drawings};
  \CCScatTwelve{Computing methodologies}{Computer graphics}{Rendering}{Ray tracing};
}

\begin{document}


\firstsection{Introduction}

\maketitle

Graph drawing is concerned with finding layouts for graphs and networks while
adhering to particular aesthetic criteria~\cite{dibattista:2000,purchase:2002}.
These can, for example, be minimal edge crossings, grouping by connected
components or clusters, and obtaining a uniform edge length. Force-directed
algorithms~\cite{eades:1984,kamada:1989} associate forces with the vertices and
edges and iteratively apply those to the layout until equilibrium is reached
and the layout becomes stationary.

\emph{Spring embedders}, as one representative of force-directed algorithms,
iteratively apply repulsive and attractive forces to the graph layout. The
repulsive force computation phase requires $O(|V|^2)$ time over the graph's
vertex set $V$. This phase can be optimized using data structures like grids or
quadtrees, as the mutually applied forces effectively only affect vertices
within a certain radius.

In this paper, we show how the task of finding all vertices within a given
radius can also be formulated as a ray tracing problem. This approach does not
only create a simpler solution by leaving the problem of efficient data
structure construction to the API, but also allows for leveraging
hardware-accelerated NVIDIA RTX ray tracing cores (RT cores).

\section{Background and Prior Work}
In the following, we provide background and discuss related work on
force-directed graph drawing algorithms. We also give an introduction
to NVIDIA RTX and prior work.

\subsection{Force-directed graph drawing}
We consider graphs $G = (V,E)$ with vertex set $V$ and edge set $E$. Each $v
\in V$ has a position $p(v) \in \mathbb{R}^2$. Edges $e \in E = \{u,v\}$, with
$u,v \in V$, are undirected and unweighted.
The Fruchterman-Reingold (FR) algorithm~\cite{fruchterman:1991} (see
Alg.~\ref{algo:fruchtermanreingold}) calculates the dispersion to displace
each vertex based on the forces. A dampening factor is used to slow down the
forces with an increasing number of iterations. Repulsive forces are computed
for each pair of vertices $(u,v) \in V$.
Attractive forces only affect those pairs that are connected by
an edge. The following force functions are used:
\begin{equation}
F_{rep}(\Delta,k) = \frac{\Delta}{|\Delta|} \cdot \frac{k^2}{|\Delta|}
\end{equation}
and
\begin{equation}
F_{att}(\Delta,k) = \frac{\Delta}{|\Delta|} \cdot \frac{\Delta^2}{|k|},
\end{equation}
where $\Delta = p(v) - p(u)$ is the vector between the two vertices acting
forces upon each other. $k$ is computed as $\sqrt{A/|V|}$, where $A$ is
the area of the axis-aligned bounding rectangle of $V$.
\begin{DIFnomarkup}
\begin{algorithm}[t]
\begin{algorithmic}
\Procedure{SpringEmbedder}{$G(V,E)$,$Iterations$,$k$}
    \For{i := 1 \textbf{to} $Iterations$}
        \State $D \gets |V|$ \Comment{ dispersion to displace vertices }

        \ForAll{$v \in V$} \Comment{ calculate repulsive forces (V x V) }
            \State $D(v) := 0$
            \ForAll{$u\in V$}
                \State $D(v) := D(v) + F_{rep}(p(v)-p(u),k)$
            \EndFor
        \EndFor

        \ForAll{$e \in E$} \Comment{ calculate attractive forces }
            \State $D(v) := D(v) - F_{att}(p(v)-p(u),k)$
            \State $D(u) := D(u) + F_{att}(p(u)-p(v),k)$
        \EndFor

        \ForAll{$v \in V$} \Comment{ displace vertices according to forces }
            \State \Call{Displace}{$v$,$D(v)$,$t$} \Comment{ $t$ is a dampening factor }
        \EndFor

        \State $t := $ \Call{cool}{$t$} \Comment{ Decrease dampening factor }
    \EndFor
\EndProcedure
\end{algorithmic}
\caption{\label{algo:fruchtermanreingold}
Fruchterman-Reingold spring embedder algorithm.
}
\end{algorithm}
\end{DIFnomarkup}

As the complexity of the first nested for loop per iteration is
$O(|V|^2)$, and by observing that the pairwise forces diminish with increasing
distance between vertices, the authors propose to adapt the computation of the
repulsive force using:
\begin{equation}
F_{rep}(\Delta,k) = \frac{\Delta}{|\Delta|} \cdot \frac{k^2}{|\Delta|} u(2k - |\Delta|),
\label{eq:modified}
\end{equation}
where $u(x)$ is $1$ if $x>0$ and $0$ otherwise.
With that, only vertices inside a radius $2k$ will have a non-zero
contribution, which in turn allows for employing acceleration data structures
to focus computations on only vertices within the neighborhood of $p(v)$.

The FR algorithm is a good match for GPUs as the three
phases---repulsive force computation, attractive force computation, and vertex
displacement---are highly parallel. The most apparent parallelization described
by Klapka and Slaby~\cite{klapka:2016} devotes one GPU kernel to each phase.
The outer dimension of the nested for-loop over $v \in V$
is executed in parallel, but each GPU thread runs the full inner loop over $u \in
V$ in Alg.~\ref{algo:fruchtermanreingold}. This reduces the time complexity to
$\Theta(|V|)$, whereas the work complexity remains $\Theta(|V|^2)$.
Force-directed algorithms---and in general graph drawing algorithms based on
nearest neighbor search---lend themselves well to massive parallelization on
distributed systems~\cite{hinge:2017,arleo:2019} or on many-core systems and
GPUs~\cite{gumerov:2008,panagiotidis:2015,uher:2016}.

Gajdo\v{s} et al.~\cite{gajdos:2016} accelerate the repulsive force computation
phase by initially sorting the $v \in V$ on a Morton curve. This order is
subdivided into individual blocks to be processed in parallel in separate
CUDA kernels. However, this process is inaccurate, as forces will only
affect vertices from the same block. The authors try to account for
that by randomly jittering vertex positions so that
some of them spill over to neighboring blocks. Mi et al.~\cite{mi:2016} use a
similar approximation but motivate that by imbalances originating from the
multi-level approach described in~\cite{hachul:2005} that they use in
combination with FR. Our approach does not use
approximations but is equivalent to the FR algorithm using the grid
optimization that was proposed in the original work.

General nearest neighbor queries have been accelerated on the GPU with
\textit{k}-d trees, as in the work of Hu et al.~\cite{hu:2015} and by Wehr and
Radkowski~\cite{wehr:2018}.
For dense graphs with $O(|E|) = O(|V|^2)$, the attractive force phase can
also become a bottleneck. The works by Brandes and Pich~\cite{brandes:2008}
and by Gove~\cite{gove:2019a} propose to choose only a
subset of $E$ using sampling to compute the attractive forces. Gove also
suggests using sampling for the graph's vertex set $V$ to improve
the complexity of the repulsive force phase~\cite{gove:2019b}. Other modifications
to the stress model exist. The COAST
algorithm by Ganser et al.~\cite{ganser:2013} extends force-directed algorithms
to support given, non-uniform edge lengths. They reformulate the stress function
based on those edge lengths so that it can be solved using semi-definite programming.
The maxent-stress model by Ganser et al.~\cite{ganser:2013b} initially solves
the model only for the edge lengths and later resolves the remaining degrees of freedom
via an entropy maximization model. The repulsive force computation in this work
is based on the classical N-body model by Barnes and Hut~\cite{barnes:1986} and
uses a quadtree data structure for the all-pairs comparison.
Hachul and J{\"u}nger~\cite{hachul:2007} gave a survey of force-directed
algorithms for large graphs.
For a general overview of force-directed graph
drawing algorithms, we refer the reader to the
book chapter~\cite{kobourov:2014} by Kobourov.

\subsection{RTX ray tracing}
NVIDIA RTX APIs allow the user to test for intersections of rays and arbitrary
geometric primitives. This technique is often used to generate raster images.
Here, Bounding volume hierarchies (BVHs) help reduce the complexity of this
test, which is otherwise proportional to the number of rays times the number of
primitives.
The user supplies a \emph{bounds program} so that RTX can generate axis-aligned
bounding boxes (AABBs) for the user geometry and \emph{build} a BVH. Now, a
\emph{ray generation program} can be executed on the GPU's programmable shader
cores that will \emph{trace} rays through the BVH using an API call. In the
\emph{intersection program}, called when rays hit the AABBs, the user can
test for and potentially report an intersection with the geometry. A reported
intersection will then be available in potential \emph{closest-hit} or
\emph{any-hit}. RTX GPUs perform BVH traversal in hardware. When RTX calls an
intersection program, hardware traversal is interrupted and a context switch
occurs that switches execution to the shader cores.

RTX was recently used to accelerate visualization algorithms like direct volume
rendering~\cite{morrical:2019} or glyph
rendering~\cite{zellmann:2020b}. RT cores have, however,
also been used for non-rendering applications, such as the point location
method on tetrahedral elements presented by Wald et al.~\cite{wald:2019b}.

\section{Method Overview}
\begin{figure*}[t]
  \centering 
  \includegraphics[width=\linewidth]{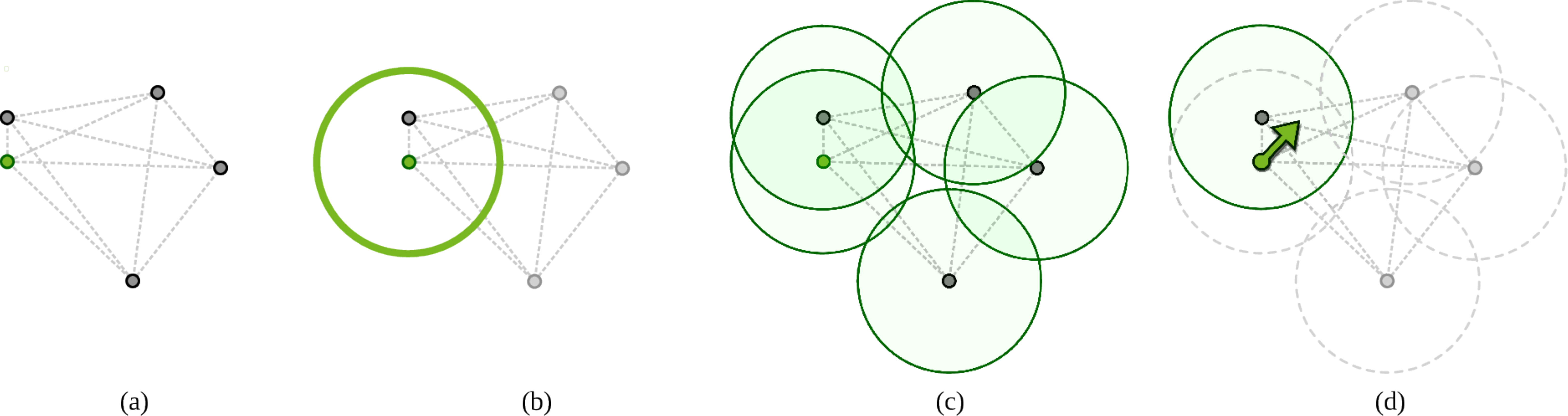}
 \caption{Mapping nearest neighbor queries to ray tracing queries. (a) The
 $K_5{:}~10$ graph; we are interested in the repulsive forces acted upon the
 green vertex by all the other vertices. (b) Nearest neighbor queries
 are performed by gathering the vertices inside a circle
 around the green vertex. (c) With a ray tracing query,
 instead of expanding a circle around the vertex of interest, we expand
 circles \emph{around all vertices}. (d) We trace an \emph{epsilon ray} (green arrow) originating
 at the green vertex' position and with infinitesimal length against the
 circles' geometry. Every circle that overlaps the ray origin---except the
 circle belonging to the vertex of interest itself---contributes to the force
 on the green vertex.}
 \label{fig:overview}
  \vspace{-1em}
\end{figure*}
We propose to reformulate the FR algorithm as a ray tracing problem. That way,
we can use an RTX BVH to accelerate the nearest neighbor query during the
repulsive force computation phase. The queries and data structures used
by the two algorithms differ substantially: force-directed algorithms use
spatial subdivision data structures, whereas RTX uses object subdivision.
Nearest neighbor queries do not directly map to the ray / primitive
intersection query supported by RTX. However, we present a mapping from one
approach to the other and demonstrate its effectiveness using an FR implementation
with the CUDA GPU programming interface.

\subsection{Mapping the force-directed graph drawing problem to a ray tracing problem}
We present a high-level overview of our approach in \autoref{fig:overview}. A
nearest neighbor query can be performed by expanding a circle around
the position $p(v)$ of the vertex $v \in V$ that we are interested in
and \emph{gathering} all $u \in V, u \neq v$ inside that circle.
To compute forces, we would perform that search query for all $v
\in V$ and would integrate the accumulation of the forces directly into the
query.

By observing that the circle we expand around $v$ always has a radius $2k$, we
can \emph{reverse} the problem: instead of expanding a circle around $v$, we
instead expand circles around \emph{all} $v \in V$.  We then trace an
\emph{epsilon ray} with infinitesimal length and origin at $p(v)$ against this
set of circles and accumulate the forces whenever $p(v)$ is inside the circle
associated with $u \in V$, given that $u \neq v$. The intersection routine of
the ray tracer only has to compute the length of the vector between the ray
origin and the center of the circle and report an intersection whenever that
length is less than $2k$. Geometrically, one can think of this as splatting,
where the splats whose footprints overlap $p(v)$ act a repulsive force upon
$v$.

The runtime complexity of the repulsive force computation phase using nearest
neighbor queries can be reduced from $\Theta(|V|^2)$ to $\Theta(|V|\log(|V|))$
using spatial indices like quadtrees~\cite{hachul:2005} or binary space
partitioning trees~\cite{lauther:2007} built over $V$. The spatial index
would have to be rebuilt on each iteration. Likewise, the ray tracing query
complexity can be reduced in the same manner using a BVH.

\subsection{Implementation with CUDA and OptiX 7}
\label{sec:implementation}
We implemented the FR algorithm with CUDA. We use separate
CUDA kernels for the repulsive and attractive forces and for the vertex
dispersion phase. Those kernels are called sequentially in a loop over
all iterations. The dispersion that is computed during the force
phases is stored and updated in a global GPU array.

The parallel attractive force phase uses atomic operations to update the
dispersion array. The repulsive phase is implemented using OptiX~7 and the
OptiX Wrapper Library (OWL)~\cite{wald:2020}. Since the number of vertices will
never change, we use a global, fixed-size GPU array for the 2-d positions that
is shared between CUDA kernels and OptiX programs. Initial vertex placement is
at random and in a square. RTX does not support 2-d primitives, so that we
construct the BVH from discs with infinitesimal thickness.

The ray generation program spawns one infinitesimal ray per vertex $v$
originating at $p(v)$; we again account for RTX being a 3-d API by setting the
z coordinates of the ray origin and direction vector to $0$ and $1$,
respectively. In this way, we can directly accumulate the dispersion inside
the intersection program and do not even have to \emph{report} an intersection
that would otherwise be passed along to a potential closest-hit or any-hit
program.

\section{Evaluation}
\label{sec:evaluation}

\begin{table*}[tb]
\setlength\tabcolsep{1.5pt}
  \caption{\label{tab:datasets}
    Statistics and average execution times on different GPUs. We use three
    artificial graphs with different connectivity and edge degrees, and
    a twitter feed graph. $c \in C$ denote connected components. Execution
    times reported are per full iteration including all phases.
  \vspace{-1em}
    }
  \scriptsize%
	\centering%
  \begin{tabu}{c|c|c|c}
    \toprule

    \includegraphics[width=1.68in,clip,keepaspectratio]{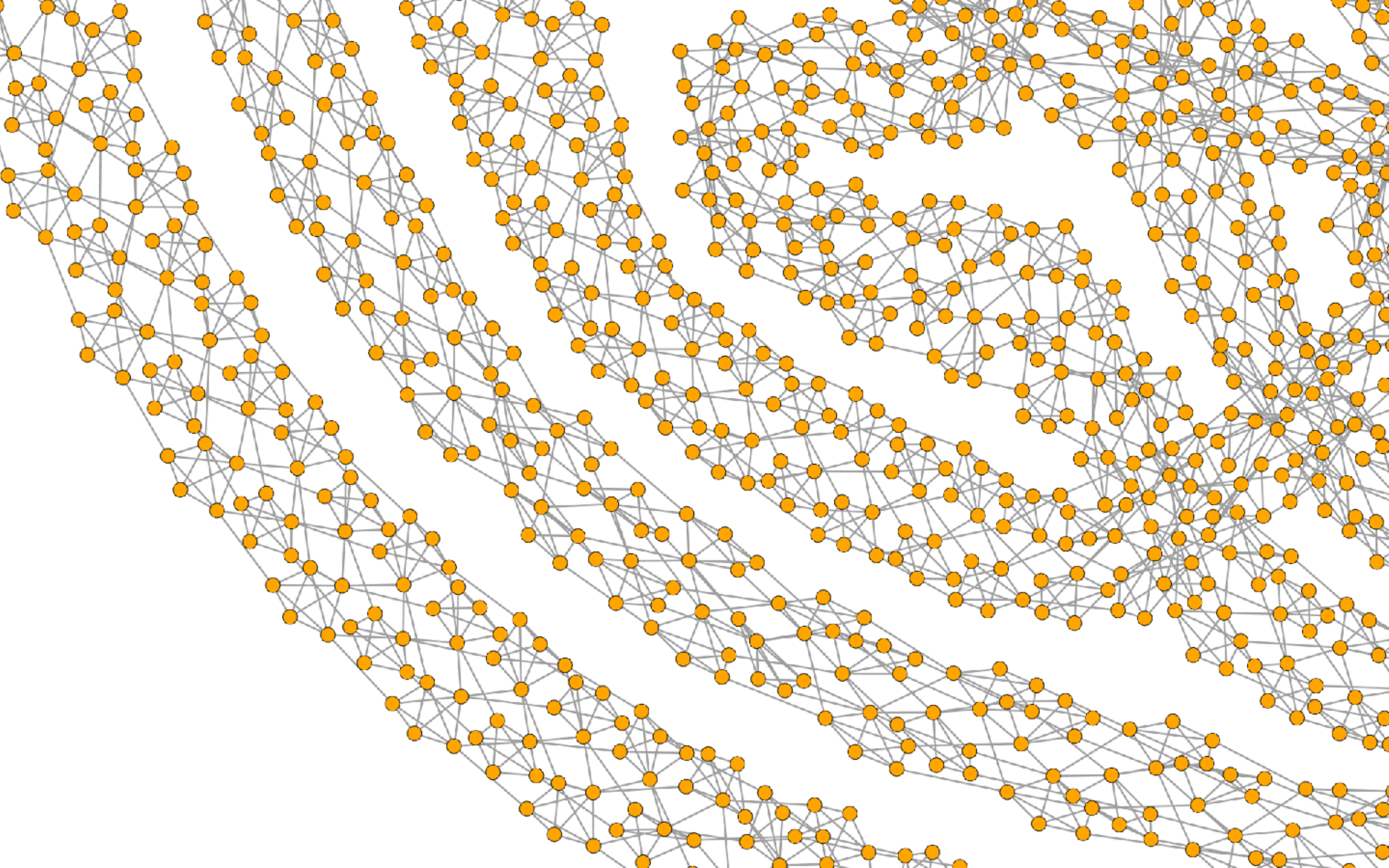} &
    \includegraphics[width=1.68in,clip,keepaspectratio]{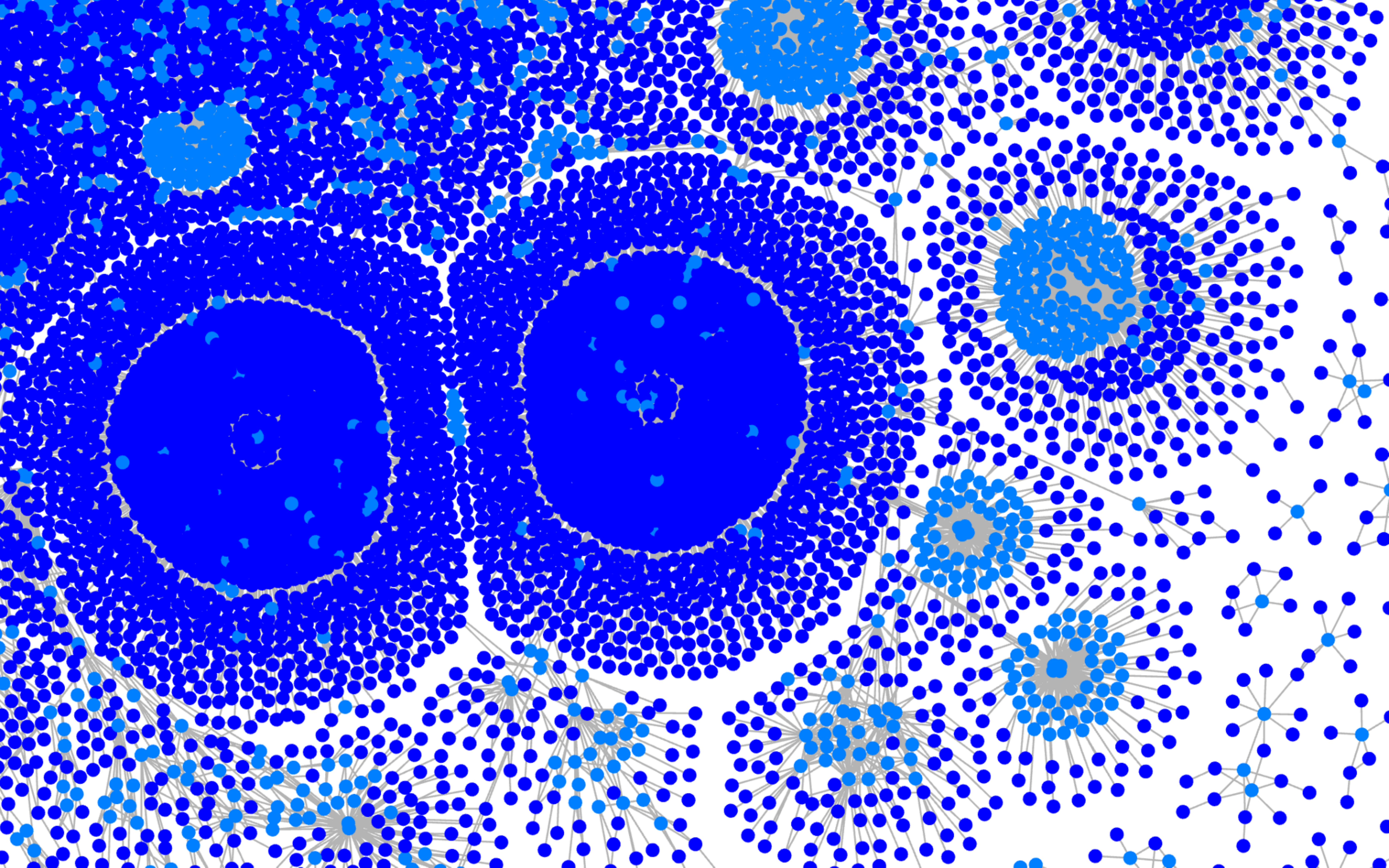} &
    \includegraphics[width=1.68in,clip,keepaspectratio]{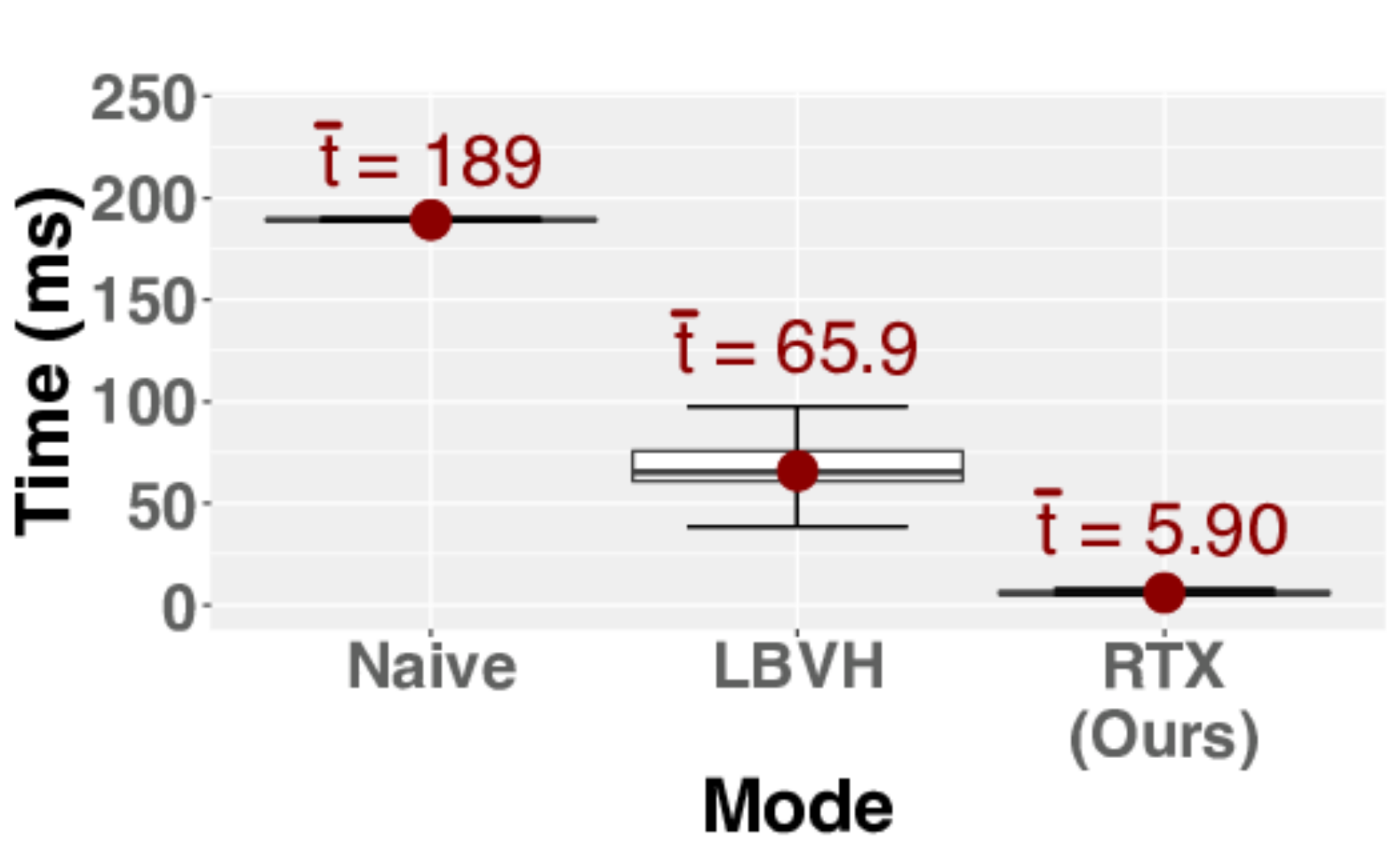} &
    \includegraphics[width=1.68in,clip,keepaspectratio]{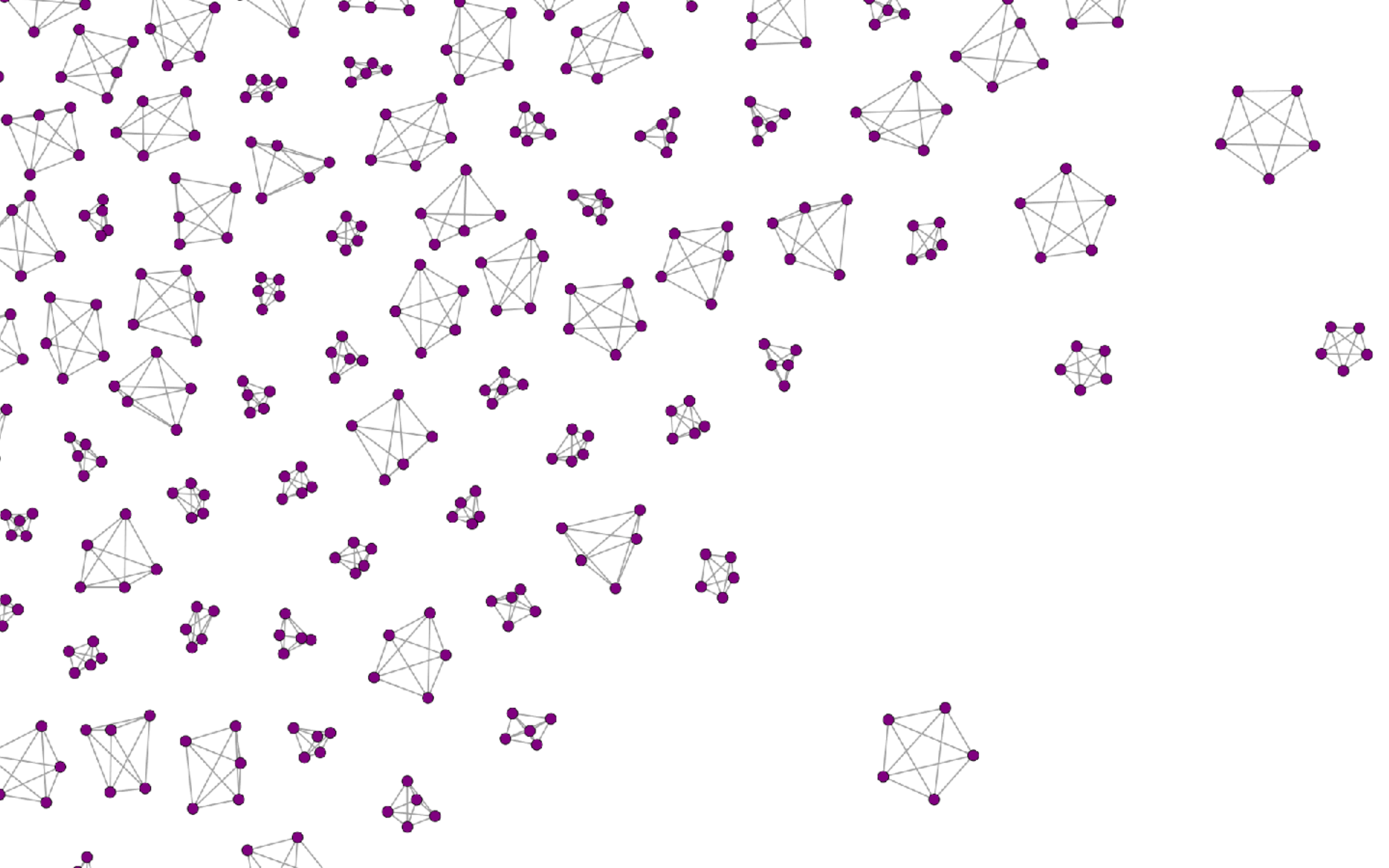}\\

    \midrule

    $5K \times K_5{:}~10$ (connected)       &   Twitter                                  & Binary Tree (Depth=16)                     & $50K \times K_5{:}~10$ (unconnected)\\
    $|V|$: $25K$, $|E|$: $69K$, $|C|$: $1$    &   $|V|$: $68K$, $|E|$: $101K$, $|C|$: $3K$   & $|V|$: $131K$, $|E|$: $131K$, $|C|$: $1$    & $|V|$: $250K$, $|E|$: $500K$, $|C|$: $50K$  \\

        Min./max./$\varnothing$ Vert. Degree: $4/8/6$ &
        Min./max./$\varnothing$ Vert. Degree: $1/810/3$ &
        Min./max./$\varnothing$ Vert. Degree: $1/3/2$ &
        Min./max./$\varnothing$ Vert. Degree: $4/4/4$ \\
        Min./max./$\varnothing$ Vert's / $c$: $25K$ (all)&
        Min./max./$\varnothing$ Vert's / $c$: $2/44K/20$&
        Min./max./$\varnothing$ Vert's / $c$: $131K$ (all)&
        Min./max./$\varnothing$ Vert's / $c$: $5$ (all)\\
    \midrule
    \includegraphics[height=2.0in,clip,keepaspectratio]{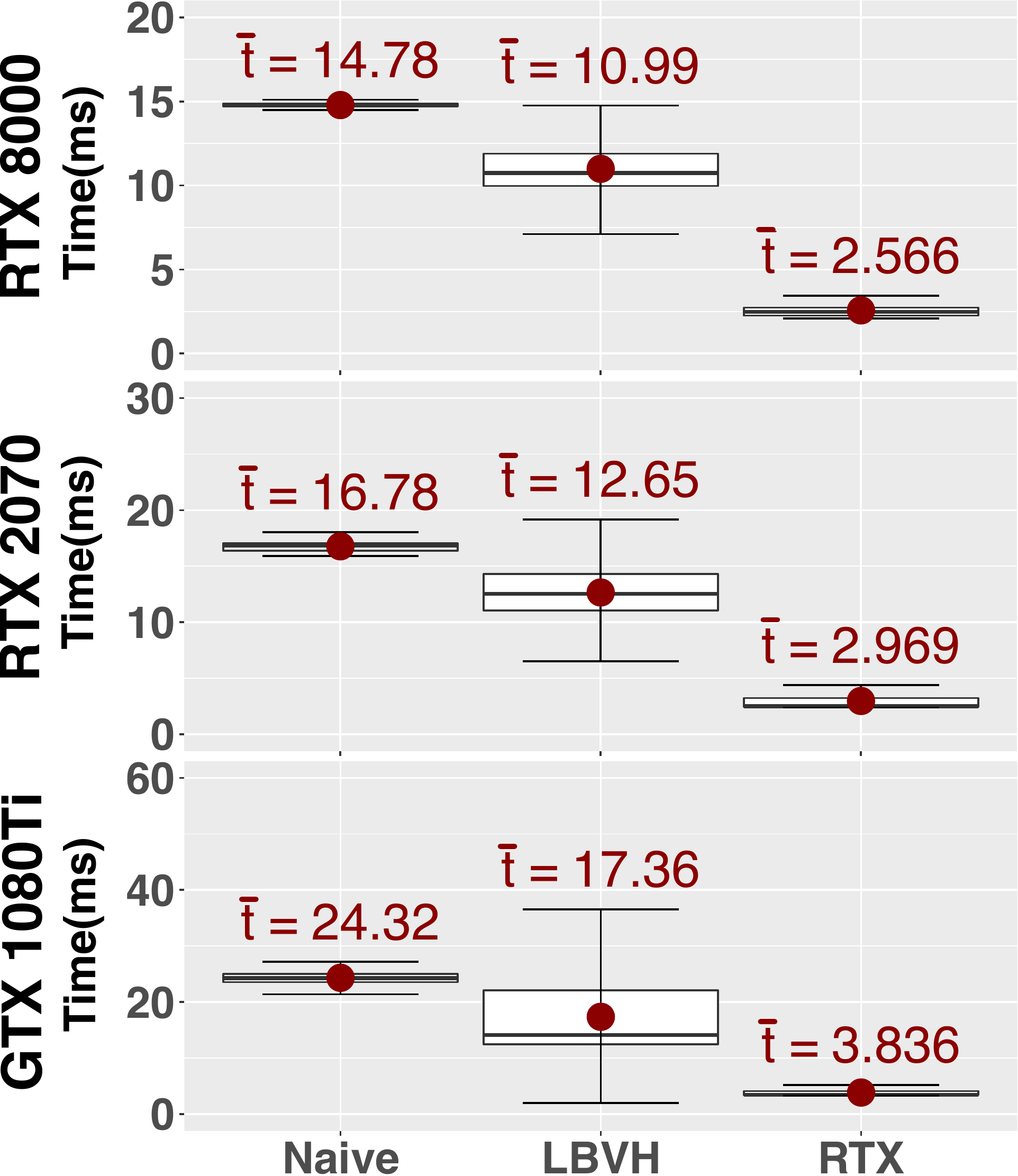} &
    \includegraphics[height=2.0in,clip,keepaspectratio]{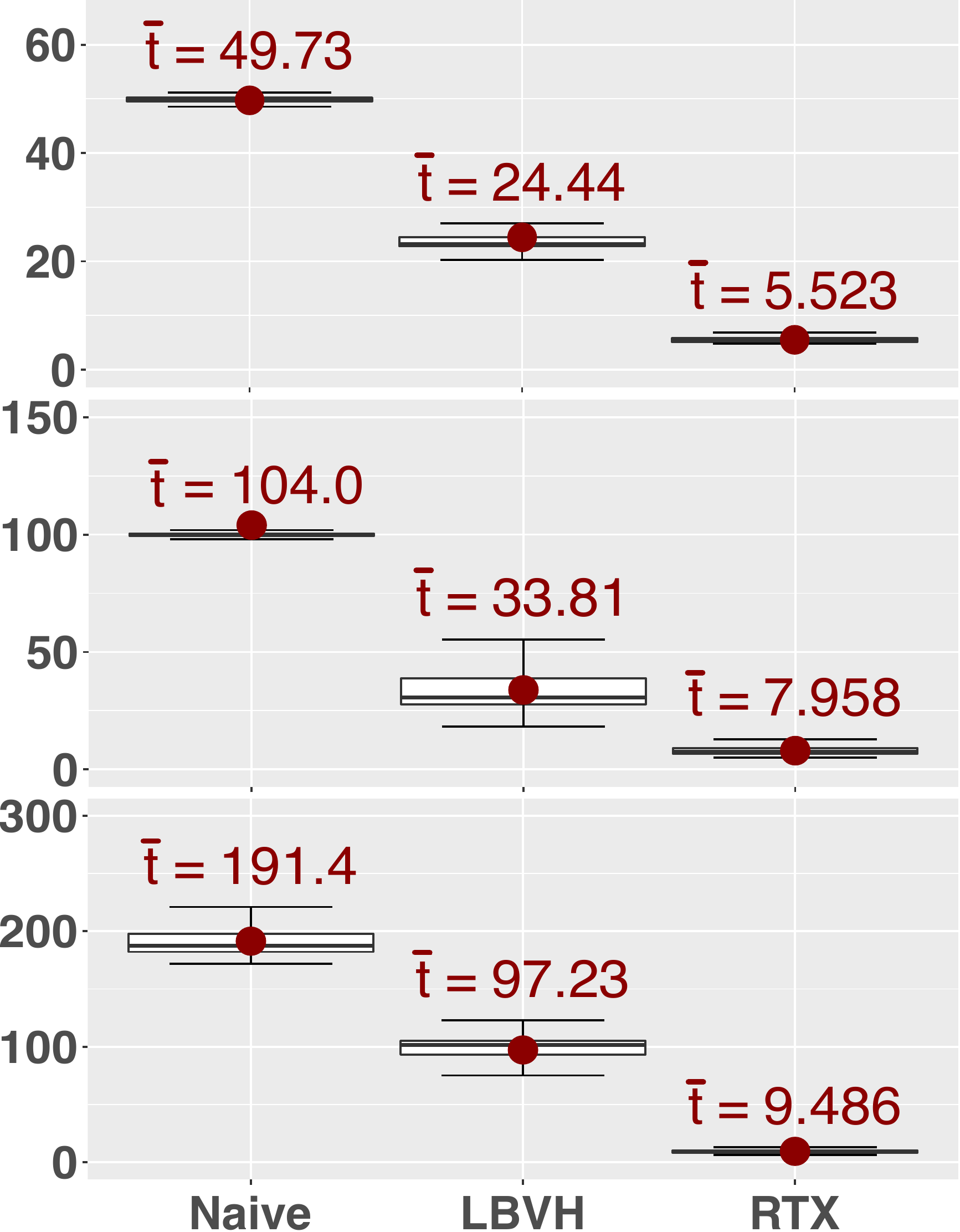} &
    \includegraphics[height=2.0in,clip,keepaspectratio]{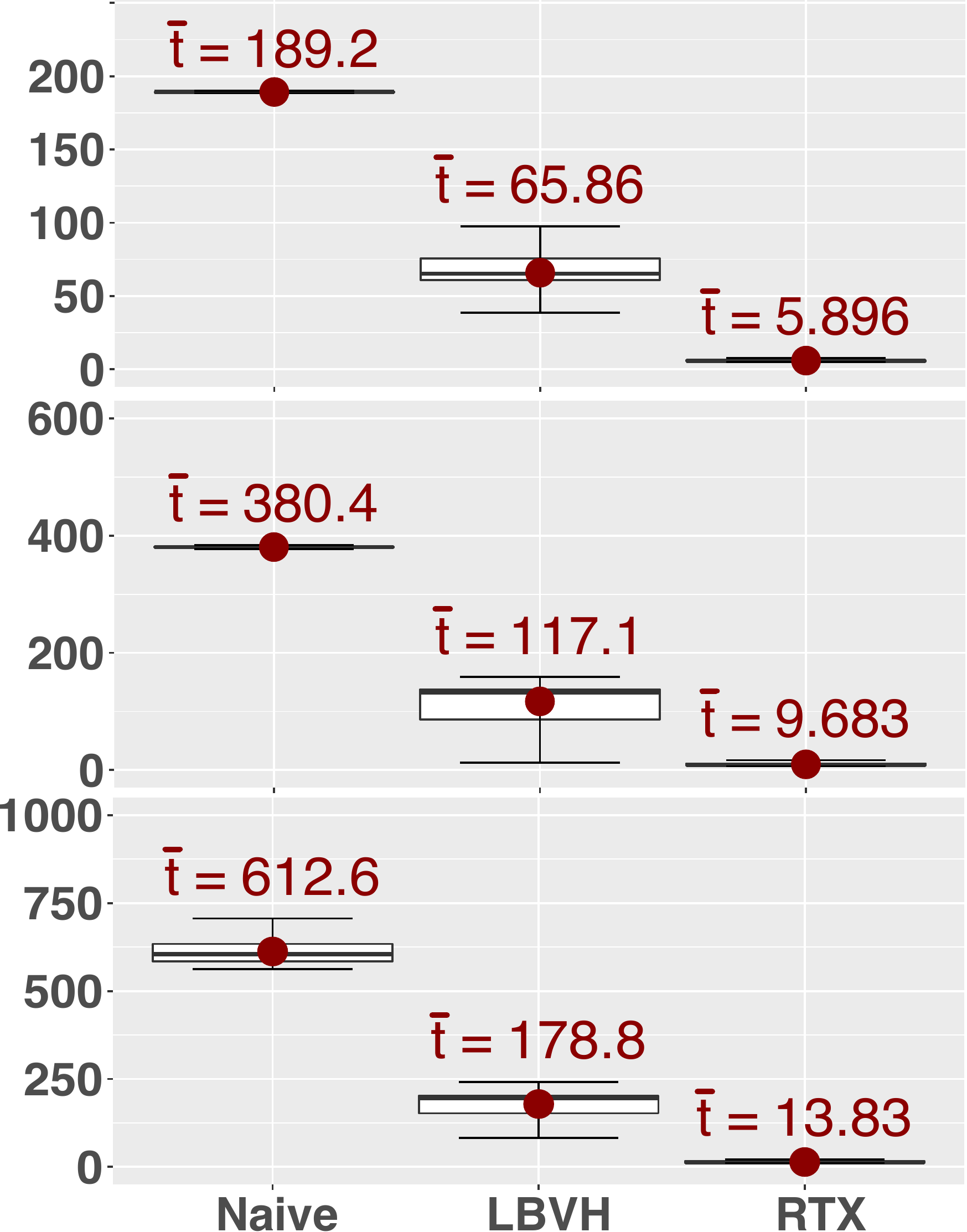} &
    \includegraphics[height=2.0in,clip,keepaspectratio]{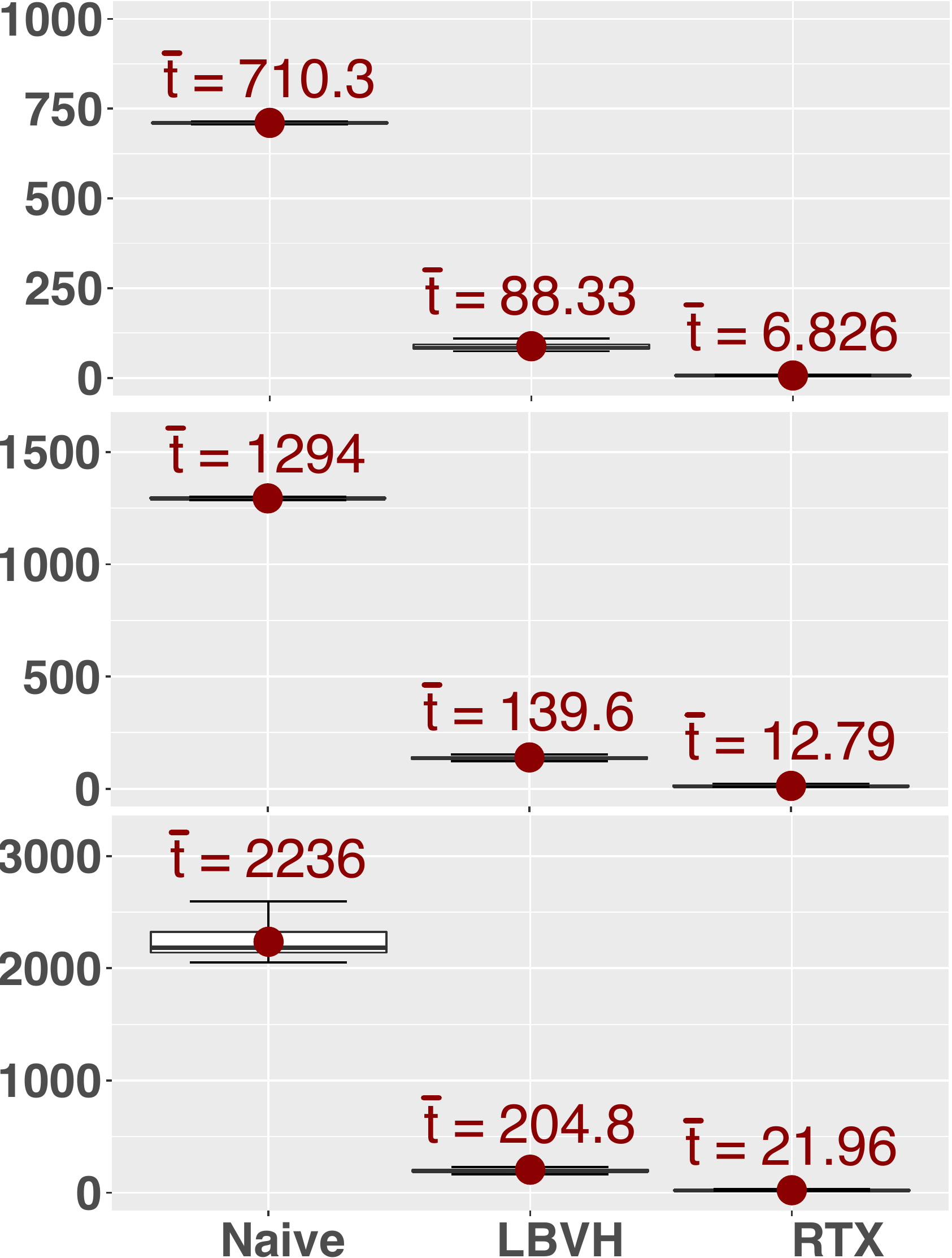}\\
    \bottomrule
  \end{tabu}
\end{table*}

\begin{table}[tb]
\setlength\tabcolsep{1.5pt}
  \caption{\label{tab:runtimeperstage}
  Acceleration data structure statistics on RTX~8000, for the repulsive
  force computation phases. Execution times per iteration are given in
  milliseconds and the ratio of build vs. traversal times in percent. We also
  report total BVH memory consumption in MB.
    }
  \scriptsize%
	\centering%

    \setlength\tabcolsep{1pt}
  \begin{tabu}{c||c||c||c|c||c|c}
  \toprule
    Data Set   &   Mode    &   Mem     &Build   &   Traversal     &     $\Sigma~F_{rep}$  &   Speedup\\
  \midrule
    $5K \times K_5{:}~10$
               & LBVH      &   1.53    & 0.92   (8.37\%)   & 10.0  (91.6\%)         &   10.9                &                       \\
   (connected) & RTX       &   1.18    & 1.16   (45.5\%)   & 1.39  (54.5\%)         &   2.55                & \textbf{4.27$\times$} \\
  \midrule
    Twitter
               & LBVH      &   4.16    & 1.94   (7.94\%)   & 22.5  (92.1\%)         &   24.4            &                       \\
        ~      & RTX       &   3.22    & 2.18   (39.7\%)   & 3.31  (60.3\%)         &   5.49            & \textbf{4.44$\times$} \\
  \midrule
    Binary Tree
               & LBVH      &   8.00    & 2.53   (3.84\%)   & 63.3  (96.2\%)         &   65.8            &                       \\
    (Depth=16) & RTX       &   6.19    & 2.36   (40.3\%)   & 3.50  (59.7\%)         &   5.87            & \textbf{11.2$\times$} \\
  \midrule
    $50K \times K_5{:}~10$
               & LBVH      &   15.3    & 2.87   (3.26\%)   & 85.4  (96.7\%)         &   88.3            &                       \\
 (unconnected) & RTX       &   11.8    & 2.82   (41.6\%)   & 3.95  (58.4\%)         &   6.77            & \textbf{13.0$\times$} \\
  \bottomrule
  \end{tabu}


  \end{table}
For a comparison with a fairly optimized, GPU-based nearest neighbor query, we
use a 2-d spatial data structure based on the LBVH
algorithm~\cite{lauterbach:2009,zellmann:2019}.
As the vertices have no area, we obtain
a 2-d BSP tree with axis-aligned split planes
that subdivide parent nodes into two same-sized halves (\emph{middle split}).
With the restriction being relaxed that two split planes need to be placed at
once, we should outperform the commonly used grid or quadtree
implementations~\cite{chimani:2014,gove:2019b}. Using Karras' construction
algorithm~\cite{karras:2012}, the build complexity is $O(n)$ in the
number of primitives. Our motivation to use a data structure with superior
construction performance is that is must be rebuild after each
iteration. We use a full traversal stack in local GPU memory and perform
nearest neighbor queries by gathering all vertices within a $2k$ radius around
the current vertex position at the leaves. We have a slight advantage over RTX
as our data structure is tailored for 2-d. At the same time we
cannot possibly optimize our data structure in the same way that NVIDIA
probably has done with RTX, and neither that this is our goal with this
comparison.

Note that the LBVH and RTX implementations and grid-based FR result in
identical graph layouts. In comparison to state-of-the-art implementations in
graph drawing libraries such as OGDF~\cite{chimani:2014}, Tulip~\cite{Auber04},
or Gephi~\cite{ICWSM09154}---all of which provide sequential CPU
implementations of FR---both our RTX and LBVH solutions are magnitudes faster.
In order to put both our GPU results into perspective, we also implemented the
naive GPU parallelization from~\cite{klapka:2016} over just the outer loop
of the repulsive force phase.

We report execution times for the four data sets depicted
in~\autoref{tab:datasets}. Two artificial data sets
consist of many fully connected $K_5{:}~10$ graphs (five vertices, ten edges).
In one case we use $5K$ of those and sequentially connect pairs of
them with a single edge. In the second case we use $50K$ of them as individual
connected components. We also test using a complete binary tree with depth $16$,
as well as the graph representing twitter feed data that is also
depicted in~\autoref{fig:teaser}. For the results reported in
\autoref{tab:datasets} we used an NVIDIA GTX~1080~Ti (no RT cores), an RTX~2070,
and a Quadro RTX~8000. The scalability study from \autoref{fig:scale}
and the evaluation of the repulsive phase in \autoref{tab:runtimeperstage}
were conducted solely on the Quadro GPU.

\section{Discussion}
\begin{figure}[tb]
  \centering
  \includegraphics[height=3.21cm,keepaspectratio]{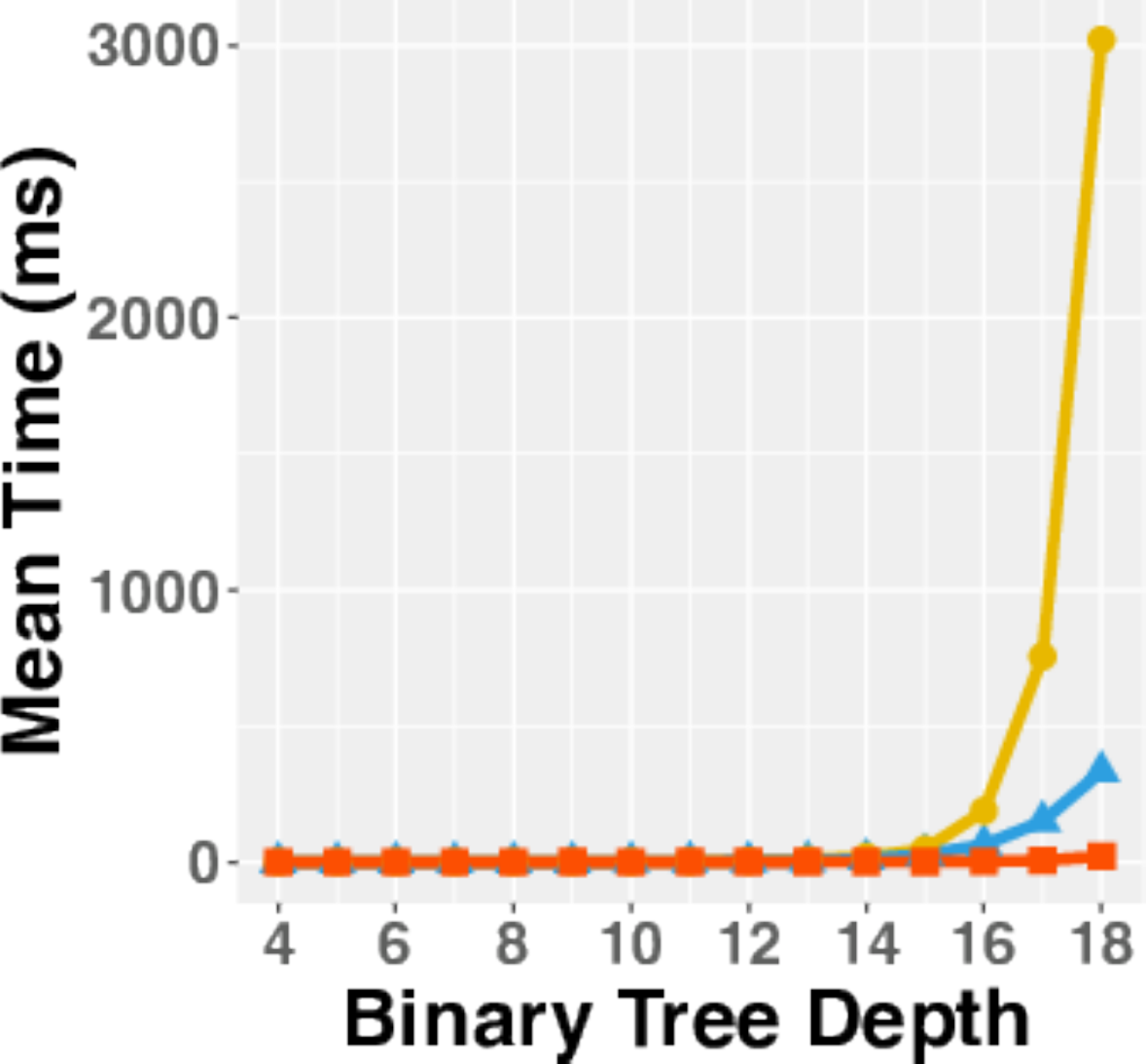}
  \includegraphics[height=3.21cm,keepaspectratio]{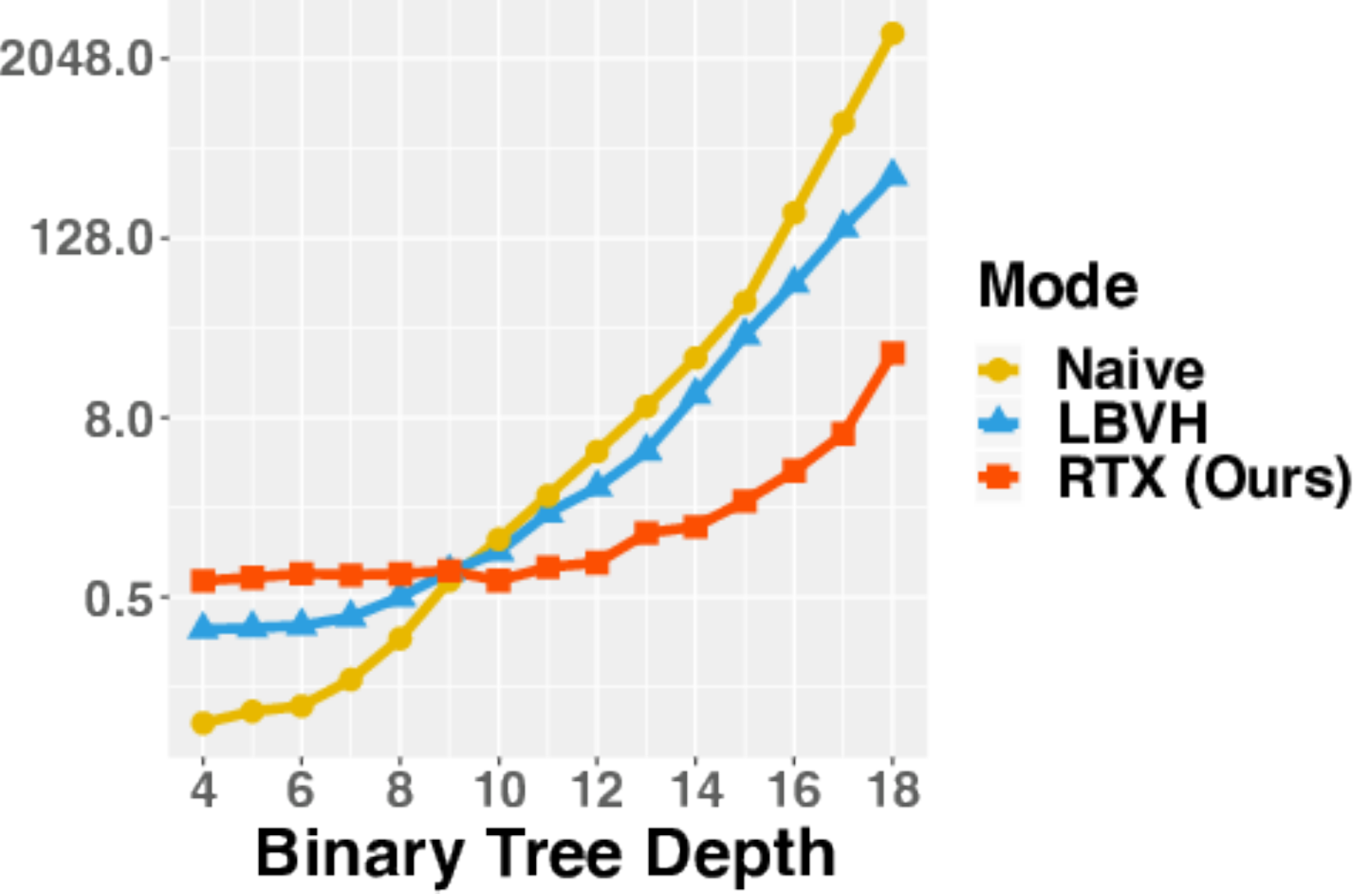}
  \caption{\label{fig:scale}
  Scalability study where we build complete binary trees with
  depth $D=4,5,\dots,18$. Left: linear scale, right: logarithmic scale.
  We report mean times for only the repulsive force phase.
  }
\end{figure}

Our evaluation suggests speedups of $4\times$ to $13\times$ over LBVH.
From the difference between the mean iteration times in \autoref{tab:datasets}
and the mean times for only the repulsive phase in \autoref{tab:runtimeperstage}
we see that the algorithm is dominated by the latter. The other phases plus overhead
account for less than $1~\%$ of the execution time.
While \autoref{fig:scale} shows that our method's performance overhead for
small graphs can be neglected---because it is on the order of about 1 ms—--we
observe dramatic speedups that increase asymptotically with $|V|$.

Interestingly, we see about the same \emph{relative} speedups
on the GeForce GTX GPU and on the RTX 2070 GPU with hardware acceleration. At
the same time, we observe that the \emph{absolute} runtimes differ
substantially, which we cannot intuitively explain, as neither the peak performance in FLOPS, nor the memory performance
of the two GPUs, differ that much. Profiling our handwritten CUDA nearest
neighbor query, we find tree traversal to be limited by L2
cache hit rate, which is about $20~\%$. For RTX, such an
analysis is impossible and we can only speculate about the results.
It is conceivable that the RTX BVH has an optimized memory layout
such as the one by Ylitie et al.~\cite{ylitie:2017}.
Assuming that we are bound by memory access latency, the speedups we
observe might stem from better utilization of the GPU's memory subsystem rather
than hardware acceleration.
Switching between hardware and software execution on RTX GPUs incurs an
expensive context switch. Hardware traversal is interrupted
whenever the intersection program is called. For our test data sets,
we consistently found the average number of intersection program instances
called to be in the hundreds. We might see an
adversarial effect where we, on the one hand, benefit from hardware acceleration,
but on the other hand suffer from expensive context switches and that the two
effects in the end cancel. We find the speedups that we observe
reassuring, especially because using RTX lifts the burden of having to program
an optimized tree traversal algorithm for the GPU from the user.


\section{Limitations of Our Study}
We acknowledge that force-directed methods for large graphs exist that
require fewer \emph{iterations} to arrive at a converged layout and outperform
FR by far in this regard~\cite{hachul:2007} and are often based
on multilevel optimizations~\cite{valejo:2020}. We chose FR as a most
\emph{simple} force-directed algorithm to reason about the speedup and
practicability of our approach.
Algorithms that perform a nearest neighbor search to compute
forces will generally benefit from the proposed techniques. The Fast Multipole Multilevel
Method ($FM^3$) \cite{hachul:2005b} employs such a nearest neighbor search and
uses a coarsening phase in-between iterations.
Similar to our method, the GPU multipole algorithm by
Godiyal et al.~\cite{godiyal:2009} employs a \textit{k}-d tree that is
rebuilt per iteration, uses stackless traversal, and would
likely benefit from RTX. The GRIP method by
Gajer and Kobourov~\cite{gayer:2001} employs a refinement phase that \emph{uses}
FR to compute local displacement vectors.
Although we assume that our approach will \emph{complement} state-of-the-art
algorithms with better convergence rates, a thorough comparison is outside of
this paper's scope and presents a compelling direction for future work.

\section{Conclusions}
We presented a GPU-based optimization to the force-directed Fruchterman-Reingold
graph drawing algorithm by mapping the nearest neighbor query performed during
the repulsive force computation phase to a ray tracing problem that can be
solved with RT core hardware. The speedup over a nearest neighbor query with a
state-of-the-art data structure that we observe is encouraging. Force-directed
algorithms lend themselves to a parallelization with GPUs.
We found that those algorithms can be optimized even further by using
RT cores and hope that our work raises awareness for this hardware feature even
outside the typical graphics and rendering communities.

\bibliographystyle{abbrv-doi}

\bibliography{egbibsample}
\end{document}